\documentclass[12pt,a4paper]{article}
\usepackage{color,graphics,amsmath,epsfig,rotating}

\usepackage{amsmath}
\usepackage{amssymb}
\usepackage{mathrsfs}
\usepackage{graphicx}
\usepackage{subfig}
\usepackage{psfrag}
\usepackage{url}
\usepackage{amssymb}
\usepackage{inputenc}
\usepackage[toc,page]{appendix}
\usepackage{hyperref}
\usepackage{comment}
 
\DeclareMathAlphabet{\mathpzc}{OT1}{pzc}{m}{it}

\usepackage{cancel} 
\begin{document}

\newcommand{\ds}{\texttt{DarkSusy}}
\newcommand{\micro}{\texttt{micrOMEGAs}}
\newcommand{\microsix}{\texttt{micrOMEGAs\,6}}
\newcommand{\py}{\texttt{PYTHIA}}
\newcommand{\HE}{\texttt{HERWIG}}
\newcommand{\msun}{{\rm M_{\odot}}}
\newcommand{\ie}{{\it i.e.} }
\newcommand{\eg}{{\it e.g.} }
\newcommand{\pbar}{${\rm \bar{p}}$}
\newcommand{\dbar}{${\rm \bar{D}}$}
\newcommand{\beq}{\begin{equation}}
\newcommand{\eeq}{\end{equation}}
\def\rsun{r_\odot}
\newcommand{\noi}{\noindent}
\def\ent{{\mathfrak{s}}}
\def\epem{e^+ e^-}
\newcommand{\calchep}{\texttt{CalcHEP}}
 \def\lnTop{\log\left(\frac{m_t^2}{Q^2}\right)}
\def\lsp{\chi}
\def\mlsp{m_{\chi}}
\def\mglu{m_{\tilde g}}
\newcommand{\SM}{\textrm{SM}}
\newcommand{\eff}{\textrm{eff}}

\newcommand{\beqn}{\begin{eqnarray}}
\newcommand{\eeqn}{\end{eqnarray}}

\newcommand{\sk}[1]{\textcolor{magenta}{#1}}

\definecolor{firebrick}{rgb}{0.7, 0.13, 0.13} 
 \newcommand{\fb}[1]{\textcolor{firebrick}{\bf{{Fawzi: #1}}}}  
 
\begin{titlepage}
\hspace{12cm}{IPPP/23/82}

\begin{center}
\vspace{1.5cm}

{\Large\bf micrOMEGAs\,6.0: N-component dark matter } \\[8mm] 

{\large   G. Alguero$^1$, G.~B\'elanger$^2$, F.~Boudjema$^2$,  S. Chakraborti$^3$, \\[2mm]
A.~Goudelis$^4$, S.~Kraml$^1$, A.~Mjallal$^2$, A.~Pukhov$^5$}\\[4mm]

{\it 
\noindent
1) Univ. Grenoble Alpes, CNRS, Grenoble INP, LPSC-IN2P3, Grenoble, France\\
2) LAPTh,  CNRS, Université Savoie Mont-Blanc, 9 Chemin de Bellevue, 74940 Annecy, France\\
3) IPPP, Physics Department, Durham University, Durham DH1 3LE, UK\\
4) Laboratoire de Physique de Clermont (UMR 6533), CNRS/IN2P3, Univ.\ Clermont Auvergne, 4 Av.\ Blaise Pascal, F-63178 Aubi\`ere Cedex, France\\
5) Skobeltsyn Inst. of Nuclear Physics, Moscow State Univ., Moscow 119992, Russia\\}


\begin{abstract}
\micro\ is a numerical code to compute dark matter (DM) observables in generic extensions of the Standard Model of particle physics. We present a new version of \micro\ that includes a generalization of the Boltzmann equations governing the DM cosmic abundance evolution which can be solved to compute the relic density of N-component DM. The direct and indirect detection rates in such scenarios take into account the relative contribution of each component such that constraints on the combined signal of all DM components can be imposed.  The co-scattering mechanism for DM production is also included, whereas the routines used  to compute the relic density of feebly interacting particles have been improved in order to take into account the effect of thermal masses of t-channel particles. Finally, the tables for the DM self-annihilation - induced photon spectra have been extended down to DM masses of 110 MeV, and they now include annihilation channels into light mesons.  
\end{abstract}

\end{center}
\end{titlepage}

\setcounter{tocdepth}{2}
\tableofcontents
\clearpage

\noindent
{\bf{PROGRAM SUMMARY}}
\normalsize
\begin{description}
\item{{\it Program title:}} micrOMEGAs6.0
\item{{\it Licensing provisions:}} GNU General Public License 3 (GPL)
\item{{\it Programming language:}} C and Fortran
\item{{\it Journal reference of previous version:}} Comput. Phys. Comm.  231 (2018)173.

\item{{\it Does the new version supersede the previous version?:}} Yes
\item{{\it Reasons for the new version:}} Previous versions of micrOMEGAs worked within the assumption that dark matter is composed of one or two components.  The new version allows for more components which can be either weekly or feebly interacting. The possibility of co-scattering is also implemented. 

\item{{\it Summary of revisions:}} This version includes new routines to compute the abundance  of multi-component dark matter that contains either weakly or feebly interacting dark matter particles in generic extensions of the Standard Model of particle physics. The co-scattering mechanism for DM production is also included. 
The routines to compute the relic density of feebly interacting particles through the freeze-in mechanism have been improved in order to take into account the effect of thermal maases of $t$-channel particles. The tables for the photon spectra resulting from  pair annihilation have been extended down to dark matter masses of 110 MeV and they now include annihilation channels into light mesons. 
\item{{\it Nature of problem (approx. 50-250 words):}}  Dark matter candidates that satisfy cosmological constraints cover a wide range of masses and interaction strengths.   Moreover, the dark sector could contain several stable neutral particles that can all contribute to dark matter.  We provide the first public code   to perform 
a precise computation of the relic density for generic extensions of the standard model with more than two component dark matter. 
\item{{\it Solution method (approx. 50-250 words):}}  We solve N Boltzmann equations treating both the cases where the dark matter components are in thermal equilibrium with the thermal bath  in the early Universe, as well as the case where the dark matter is too feebly interacting to reach equilibrium. We also include decay terms in the Boltzmann equations.
All the signals for dark matter direct and indirect detection take into account the contribution of each component to the total relic density. 
\end{description}

\section{Introduction}

Despite ample evidence for the existence of dark matter (DM) at the galactic, galaxy cluster and cosmological scales, the nature of DM remains a complete mystery. Starting from the well-motivated  hypothesis that it consists of one or more new particles in some extension of the Standard Model (SM) of particle physics, extensive experimental programs have been actively searching for DM.
In order to assess the viability of different DM models given the abundant data coming from cosmological observations,
astrophysical direct and indirect searches and collider searches for new particles, several computational tools have been developed in order to  accurately compute DM observables. In particular, most publicly available tools were developed in order to calculate the relic density of DM  under the assumption that its cosmic abundance is due to the freeze-out mechanism~\cite{Belanger:2006is, Binder:2021bmg, Gondolo:2004sc, Ambrogi:2018jqj, Arbey:2009gu}, although in recent years alternative DM generation mechanisms have been gradually incorporated in some of these tools -- most notably freeze-in~\cite{Belanger:2018mqt,Bringmann:2018lay}. These codes can be used to compute the predicted DM abundance within specific particle physics models~\cite{Arbey:2009gu} or in generic extensions of the Standard Model~\cite{Belanger:2006is,Ambrogi:2018jqj,Bringmann:2018lay} and allow, in addition, the computation of a variety of DM-related observables.

Under the assumption that DM is a new weakly interacting massive particle (WIMP), the computation of its relic density follows the formalism described in~\cite{Gondolo:1990dk} and relies on the premise that, at sufficiently early cosmic times, the DM as well as all other exotic particles were in kinetic and chemical equilibrium with the particles of the SM. Within these assumptions a single equation can be written to describe the evolution of the DM abundance and then derive the relic density~\cite{Edsjo:1997bg}. This is the case even when the ``dark sector'' (DS) contains a large number of new particles that can, depending on the spectrum, all contribute to co-annihilation processes.\footnote{Here the term ``DS'' is used to signify each set of beyond the SM particles that i) transform in the same way under a discrete symmetry that guarantees the stability of the lightest particle carrying non-trivial charge under the given symmetry and ii) are in thermal equilibrium with each other. In the following this definition will be slightly extended in order to include the case of Feebly Interacting Massive Particles, FIMPs.} There are, however, several cases in which this simplified treatment of the relic density calculation needs to be extended. 

Firstly, when there are many DSs, potentially each containing a stable particle, the possibility of multi-component DM opens up. In this case, an evolution equation for the abundance of each DM component is required.  This occurs, \textit{e.g.}, in models in which the charge assignments under one or more discrete symmetries lead to several stable particles. For instance, symmetries such as $Z_4$~\cite{Belanger:2012vp,Belanger:2021lwd}, $Z_5$~\cite{Belanger:2020hyh} or $Z_2\times Z_2$~\cite{Belanger:2011ww,Aoki:2012ub,Bhattacharya:2016ysw} can lead to two DM components, while a $Z_7$ symmetry can feature 3-component DM~\cite{Belanger:2022esk}. The solution for the two-component case was already included in one of the publicly available DM computational tools, namely \texttt{micrOMEGAs4.1}~\cite{Belanger:2014vza}, whereas the three (or more)-component case requires a generalisation of the corresponding framework. 

Besides, the number of DM evolution equations also needs to be extended if the chemical equilibrium condition does not hold even within a sector containing particles with the same charge under the discrete symmetry. The processes that are responsible for maintaining chemical equilibrium within a sector include decay processes, such as $\chi'\to \chi\,\SM$, scattering processes such as $\chi\chi'\to \SM\,\SM$ or inelastic scattering processes such as $ \chi\,\SM \to \chi'\,\SM'$. Here $\chi,\,\chi'$ denote any particle(s) in a DS. In the first case, chemical equilibrium is maintained if the decays of DS particles are faster than the Hubble expansion rate, $H$. In the other cases, the relevant quantity to be compared with $H$ is the thermally averaged scattering cross section times the number density ($n \left\langle \sigma v \right\rangle$).  If these conditions are not satisfied, one must solve a set of Boltzmann equations for each set of particles that are  not in chemical equilibrium.  
Moreover, the inelastic scattering processes play an important role in the determination of the DM relic density when the  standard annihilation processes that govern the freeze-out mechanism, $\chi\chi^{(')}\rightarrow \SM\,\SM$, are suppressed, for example when the couplings describing the relevant interactions are small. 
In this scenario, dubbed ``conversion driven freeze-out'' or ``co-scattering''~\cite{Garny:2017rxs, DAgnolo:2017dbv,Brummer:2019inq,Cheng:2018vaj,DAgnolo:2019zkf},
the inelastic process  generates an  effective width $\Gamma_{\chi \to \chi'} = \langle v \sigma_{\chi, SM  \to \chi', SM} \rangle n_{\rm SM} $ for the  $\chi \to \chi'$ transition in the standard model bath.

Lastly, another scenario which may require a departure from the single-evolution-equation case is the one in which some dark particles are extremely weakly (``feebly'') interacting (``FIMPs''). Such particles cannot reach thermal equilibrium with other particles in the plasma; therefore, in the conventions followed in this paper, each species of such particles will be taken to form a DS of its own. FIMPs can be produced either from the decay of bath (or other feebly coupled) particles or from scattering processes involving bath particles via the freeze-in mechanism~\cite{McDonald:2001vt,Hall:2009bx,Belanger:2018mqt}. It is generally assumed, and this is a hypothesis that we will follow throughout this work, that the FIMP initial density vanishes. Note that extremely weakly interacting particles can be also produced from the decay of another DS particle which was initially in thermal equilibrium, after it freezes-out. This scenario is known as the ``superWIMP'' mechanism~\cite{Feng:2003xh}. 

 A general solution for the relic density of multi-component DM scenarios must, therefore, consider a set of Boltzmann equations which can describe any number of WIMPs and FIMPs along with decay and scattering terms involving particles from different DSs. The general solution must also include decay and scattering terms responsible for co-scattering or DM conversion.  Note that the same approach of solving sets of Boltzmann equations can  be used both for models with more than one stable DM particle as well as for models with a single DM candidate but containing sets of particles that are not in thermal equilibrium with each other or with the SM bath. Providing a computational framework in order to study such scenarios constitutes the major upgrade of the \micro\ package that will be presented in this work.

In this paper, we describe the new functionalities of \micro, a code that allows to compute the DM relic density in generic extensions of the standard model of particle physics, both within the standard freeze-out and for the freeze-in picture. The new features include:
 
\begin{itemize}
\item{} A generalisation of the Boltzmann equations in order to include multi-component DM models, including models with multiple WIMPs, FIMPs, or both WIMP and FIMP DM candidates.
\item{} A generalisation of the Boltzmann equations in order to include the conversion-driven freeze-out mechanism (or co-scattering) as well as decay terms for unstable particles in each DS~\cite{Alguero:2022inz}.
\item{} The possibility for the user to define explicitly which sets of particles are in thermal equilibrium; a dedicated function also allows to \textit{check} whether chemical equilibrium is maintained throughout the cosmic evolution.
\item{}  In multi-component DM models, all components are taken into account when computing the direct and indirect detection rates. Moreover, these rates can be rescaled according to the relative contribution of each component to the total DM density.
\item{} For single-component DM models, the user has the additional option of including specific $2\to 3$ or $2\to 4$ processes mediated by one or two virtual particles.
\item{} An improvement in the computation of the relic density of FIMPs: the effect of the thermal mass of particles exchanged in $t$ or $u$-channel Feynman diagrams is simulated by introducing an appropriate integration cut.
\item{} The possibility to impose constraints from direct detection experiments on single- and multi-component DM models, also including non-standard assumptions about the DM velocity distribution or the nucleus form factors. 
\item{} The tables for DM-annihilation-induced $\gamma$ production, relevant for indirect detection are extended to masses below  $2$ GeV. 
\item{A computation of the free-streaming length in models with long-lived particles. This feature allows the user to estimate constraints stemming from Lyman-$\alpha$ forest observations.}
\end{itemize}

The paper is structured as follows: in Section~\ref{sec:Ncomp} we present the evolution equations for multi-component DM and their solutions as implemented in \microsix. 
In Section~\ref{sec:coscat} we summarize the conversion-driven freeze-out DM production mechanism and, again, the way it is incorporated in the code. In Section \ref{sec:singularities} we address some issues which can appear once one relies solely on zero-temperature field theory calculations when computing the DM relic abundance and which can lead to erroneous predictions. 
In Section~\ref{sec:other} we present some additional improvements that have been incorporated in \microsix\ whereas in Section~\ref{sec:limits} we describe some updates concerning the implementation of experimental limits. 
In Section \ref{sec:functions} we present and explain the new routines included in the code and in Section~\ref{sec:examples} we provide an example concerning the utilization of the code. Lastly, in Section~\ref{sec:conclusions} we conclude. 
\section{Multi-component DM} 
\label{sec:Ncomp}
We begin by discussing the most important upgrade introduced in \microsix\ with respect to previous versions of the code: the possibility to compute the DM abundance in multi-component DM models. The relevant Boltzmann equations will be presented in a manner which allows  to cover the case of multi-component DM including the possibility of conversion-driven freeze-out.

\subsection{Thermodynamics in a radiation-dominated Universe}

In order to render our discussion as self-contained as possible, let us begin by recalling some useful relations concerning the thermodynamics of the Universe and describe how some important relevant quantities can be accessed in {\tt micrOMEGAs}.  

First, the effective numbers of degrees of freedom related to the entropy ($\ent$) and energy ($\rho_R$) densities in a radiation-dominated Universe, $h_{\eff}(T)$ and $g_{\eff}(T)$ respectively, are given by
\begin{equation}
  \rho_R(T) = \frac{\pi^2}{30} g_{\eff}(T) T^4   \;\;\;  {\rm and}   \;\;\;  \ent(T)=\frac{2\pi^2}{45} h_{\eff}(T) T^3.
\end{equation}
The entropy and energy densities are related through\footnote{This equation is not valid for $T \le 1$ MeV, where photons and neutrinos have different temperatures.}
\begin{equation}
    \frac{d\rho}{dT} = T\frac{d \ent}{dT}  
    \label{eq:entropy}
\end{equation}
In  most DM models, at high temperatures (\textit{e.g.}\ close to the freeze-out temperature), only the SM particles contribute significantly to  $g_{\eff}$ and $h_{\eff}$. At the opposite end, \textit{i.e.}\ at very low temperatures ($T\ll 100$ MeV), the effective number of degrees of freedom is essentially determined by photons, neutrinos and light leptons.\footnote{The contribution of  new light degrees of freedom that can be present in certain models  is not considered here.} 
Above $\sim 100$~MeV, one has to gradually add the light mesons whereas above the temperature of the QCD phase transition, $T_{\rm QCD}$, the latter are replaced by the  quark degrees of freedom. 

Different matching schemes between the low- and high-temperature regions were studied in~\cite{Hindmarsh:2005ix}, while lattice calculations were used in~\cite{Drees:2015exa,Laine:2015kra} to improve the treatment of the temperature dependence of the entropy and energy density.  
In \micro, the default tables for $h_{\eff}$ and $g_{\eff}$ correspond to the ones presented in Ref.~\cite{Drees:2015exa} and can be found in the file \verb|Data/hgEff/DHS.thg|.
The  directory  {\tt Data/hgEff} also contains solutions described in Ref.~\cite{Laine:2015kra} (file {\tt LM.thg}),  and in Ref.~ \cite{Hindmarsh:2005ix} (files {\tt HP\_B.thg} and {\tt HP\_C.thg}), as well as the tables used in {\tt DarkSUSY} (file {\tt GG.thg}). The latter  were  used as default in  previous versions of \micro. Only the tables \verb|DHS.thg| and \verb|LM.thg| include the contribution of the Higgs boson. Besides the options that are already available in \micro, a dedicated function ({\tt loadHeffGeff}) allows the user to substitute any table. A more detailed description of this routine will be given in Section~\ref{sec:functions}. As shown in Ref.~\cite{Belanger:2013oya}, different choices for $h_{\eff}$ can induce a shift in the value of the predicted relic density, typically by a few percent. The larger shifts are expected for light DM masses, especially in case the freeze-out temperature is near the QCD transition region~\cite{Srednicki:1988ce}.

The Hubble parameter, $H(T)$, is defined via the total matter/energy density of the Universe,
\begin{equation}
   \label{eq:H}
   H= \sqrt{\frac{8\pi \rho(T)}{ 3 M_P^2}} \, ,   
   \end{equation}
where $M_P$ is the Planck mass and 
\begin{equation}
   \rho(T)=\frac{\pi^2}{30} g_{\eff}(T) T^4 + \mu_{\rm M}\frac{2\pi^2}{45} h_{\eff}(T) T^3 + \mu_{\rm DE}^4 
   \end{equation}
describes the  contributions from radiation, dark and baryonic matter ($\mu_{\rm M}$) and, lastly, dark energy ($\mu_{\rm DE}$). The values of these parameters, taken from the Particle Data Group (PDG) report \cite{ParticleDataGroup:2010dbb}, are $\mu_{\rm M}=0.519$ eV and  $\mu_{\rm DE}=2.24\;10^{-3}$ eV respectively.

Eq.~\eqref{eq:entropy}, which essentially corresponds to entropy conservation, also allows  to establish a relation between cosmic time and temperature through
\begin{equation}
  dt=-\frac{dT}{\overline{H}(T) T} 
  \end{equation}
where
\begin{equation}
\overline{H}(T) = \frac{H(T)}{\left(1+ \frac{1}{3} \frac{d\log(h_{\eff}(T))}{d\log(T)}\right)} \, .
\end{equation}
Then, the time interval during which the temperature of the Universe decreases from $T_1$ [GeV]  to $T_2$ [GeV] is obtained by integrating the Hubble rate given by Eq.~\eqref{eq:H}.
\begin{equation}
\label{eq:hubbletime}
  t= \int \limits_{T_2}^{T_1}  \frac{dT}{\overline{H}(T) T}\, . 
\end{equation}
For example, the time interval for the temperature of the Universe to drop from a sufficiently high temperature, say $10$ GeV, down to the current temperature of $2.725$ K is calculated to be $13.806$  Gyr, in agreement with the value quoted by the Particle Data Group~\cite{ParticleDataGroup:2010dbb} for the age of the Universe, namely $13.797(23)$ Gyr. 

\subsection{Evolution equations}
\label{sec:evolutioneqs}

We now move on to discuss the equations governing the time/temperature evolution of the abundance in $N$-component DM models.

\subsubsection{Standard freeze-out case}
\label{sec:evolutioneqsFO}

In order to better streamline our presentation let us, first, consider the case of models with multiple WIMP-like (\textit{i.e.}\ freezing-out) DM components. Along the lines of previous versions of {\tt micrOMEGAs} (and, arguably, a substantial number of DM models that exist in the literature), throughout our discussion we will focus on theories in which one or more discrete symmetries are imposed at the Lagrangian level, with different (sets of) particles potentially transforming differently under their direct product. All exotic particles with the same discrete symmetry transformation properties will be taken to belong to the same ``symmetry sector''. Each of these symmetry sectors (\textit{i.e.}\ each set of particles sharing the same discrete symmetry transformation properties) can be further divided into subsectors within which all particles are in chemical equilibrium. Particle species belonging to the same symmetry sector and which are in chemical equilibrium between them are taken to form a distinct ``DS''. Different DSs will henceforth be labelled with  Greek letter indices, whereas Latin indices will be used to designate different particle species within a given DS. By convention, in the following we will denote the sector consisting of all SM particles along with any new particle species with the same discrete symmetry transformation properties as the SM as sector 0. Moreover, throughout our analysis we will assume \textit{kinetic} equilibrium of all DS particles with the SM bath particles (and, hence, amongst them). Since, by definition, chemical equilibrium within each DS, say $\alpha$, holds, the number density $n_i$ for each particle species in a given DS can be obtained from the species' thermally averaged number density ($\bar{n}_i$) and the total and thermally averaged number densities of particles in that sector ($n_{\alpha}$, $\bar{n}_\alpha$) through
\begin{equation}
 n_i =  n_\alpha \frac{\bar{n}_i}{\bar{n}_\alpha} \, ,
 \end{equation}
where
\begin{equation}
\label{Neq}
\bar{n}_{\alpha}= \frac{T}{2\pi^2} \sum_{i \in \alpha} g_i m^2_i K_2(\frac{m_i}{T})
\end{equation}
and $K_2$ is the modified Bessel function of the second kind of order 2. The abundance $Y$ is related to the number density $n$ through
\begin{equation}
 Y \equiv n/\ent(T) \,,
\end{equation} 
where  $\ent(T)$ is the total entropy density. When particles are in thermal equilibrium within a given DS their equilibrium abundance is, analogously, given by 
\begin{equation}
\label{Yeq}
 \overline{Y}_{\alpha}= \bar{n}_{\alpha}/\ent(T) \,.
\end{equation}
In the multi-component DM case we will ignore effects related to the different spin-statistical distributions of particle species (Fermi-Dirac/Bose-Einstein), even though these can be quantitatively relevant for the case of FIMPs~\cite{Belanger:2018mqt}.\footnote{This approximation is adopted for reasons of computational efficiency. Note, however, that in {\tt micrOMEGAs} statistical effects can be fully accounted for in the case of single-component DM, as described in \cite{Belanger:2018mqt}.} 
The number of events of the type $a,b \to c,d$ per unit space-time volume assuming equilibrium densities for all incoming particles reads
\begin{equation} 
\label{nEvents22} 
          {\bar N}_{a,b \to c,d}= \frac{T g_a g_b }{ 8\pi^4} \int \sqrt{s}p_{ab}^2(s) K_1(\frac{\sqrt{s}}{T})  C_{ab} \sigma_{a,b\to c,d}(s) ds \,,
\end{equation}
where $C_{ab}$ is a combinatoric factor, $C_{ab}= 1/2$  if $a=b$ and  1 otherwise, $g_a$ is the number of intenal degrees of freedom and $p_{ab}$ is the momentum of the incoming particles $a$ and $b$ in their centre-of-mass frame. Throughout the paper $s$ is the usual Mandelstam variable (with $\sqrt{s}$ representing the total energy of the system). The detailed balance equation implies that
\begin{equation}
\label{balance}
           {\bar N}_{a,b \to c,d} = {\bar N}_{c,d \to a,b} 
\end{equation}
and we introduce the function 
\begin{equation}
\label{vSigmaNExp}
  \langle v \sigma_{ \alpha\beta\gamma \delta}  \rangle = \frac{1}{C_{\alpha\beta} \bar{n}_{\alpha}(T) \bar{n}_{\beta}(T)}
\sum_{\substack{a\in\alpha, b\in\beta,c\in\gamma,d\in \delta\\  {\rm if}(\alpha=\beta) a\le b;\;
{\rm if}(\gamma=\delta) c\le d }}  {\bar N}_{a,b \to c,d} \,.
\end{equation}
Under these notations, and assuming that each DS contains at most one DM candidate, the equation describing the evolution of the abundance of the $\mu$-th DM candidate only assuming $2\to2$ reactions reads 
\begin{equation}
\label{dndt}
    \frac{dn_{\mu}}{dt} = - \sum_{\alpha\le\beta;\; \gamma \le \delta} n_{\alpha}n_{\beta}\, C_{\alpha\beta}   \langle v \sigma_{ \alpha\beta\gamma \delta}  \rangle  ( \delta_{\mu\alpha} + \delta_{\mu\beta} - \delta_{\mu\gamma}
-\delta_{\mu\delta}) -3H(T)n_{\mu}    \,,
\end{equation}
where $H(T)$ is the Hubble expansion rate.

Usually, the DM evolution equations are solved in terms of the abundance $Y_\mu$. The entropy conservation equation
\begin{equation}
   \frac{d\ent}{dt}=-3H\ent
\end{equation}
allows to convert the time evolution equation into an evolution equation with respect to the entropy density $\ent$ as
\begin{equation}
  \label{eq:dYdT} 
    3H \frac{dY_{\mu}}{d\ent} = \sum_{\alpha\le\beta;\; \gamma \le \delta}
 Y_{\alpha}Y_{\beta}
C_{\alpha\beta}  \langle v \sigma_{ \alpha\beta\gamma \delta}  \rangle ( \delta_{\mu\alpha} + \delta_{\mu\beta} - \delta_{\mu\gamma}
-\delta_{\mu\delta}) \,.
\end{equation}
Writing the evolution equations in terms of the entropy density rather than in terms of the temperature allows for  a more compact notation; one can recover the corresponding equation in terms of temperature using
 \begin{equation}
  3H\frac{dY_\mu}{d\ent}= \frac{3H}{ \frac{d\ent}{dT}} \frac{d Y_\mu}{dT}=  \frac{\bar{H}T}{\ent}\frac{d Y_\mu}{dT} \,
  \label{eq:dsdT}
\end{equation}  
\\
Now, the number of collision events $\alpha\beta \rightarrow \gamma\delta$ per unit space-time volume is 
\begin{equation}
     N_{\alpha\beta\gamma\delta}=C_{\alpha\beta} \langle v\sigma_{\alpha\beta\gamma\delta} \rangle n_\alpha n_\beta \,.
\end{equation}
In the specific case where the only relevant processes are the annihilations of two DM particles of the same sector $\mu$ into bath (sector $0$) particles, Eq.~\eqref{eq:dYdT} reduces to its familiar form,
\begin{eqnarray}
  3H  \frac{dY_{\mu}}{d\ent}   &=& \langle v\sigma_{\mu\mu 00}\rangle(Y_{\mu}^2- \bar{Y}_\mu^2) \,,
\end{eqnarray} 
where we have used the detailed balance equations 
\begin{eqnarray}
\bar{N}_{\alpha,\beta \to \gamma, \delta} &=&\bar{N}_{ \gamma, \delta  \to \alpha,\beta } \,, \\
C_{\alpha\beta} \langle v\sigma_{\alpha\beta\gamma\delta} \rangle  \bar{Y}_\alpha \bar{Y}_\beta  &=& C_{\gamma\delta} \langle v\sigma_{\gamma\delta\alpha\beta} \rangle  \bar{Y}_\gamma \bar{Y}_\delta \,,
\end{eqnarray}  
and $Y_0=\bar{Y}_0$.
\\
\\
Throughout the previous discussion we have assumed that all DS particles decay rapidly. In scenarios in which the decay rate of some particle in the DS is slow, the corresponding decay processes enter explicitly the evolution equation for the DM abundance. The RHS of Eq.~\eqref{eq:dYdT} then contains additional terms of the type
\begin{equation}
\frac{1}{\ent^2(T)} \sum_{\alpha;\; \gamma \le \delta} \left( \frac{Y_{\alpha}} {\bar{Y}_{\alpha}}
- \frac{Y_\beta}{\bar{Y}_\beta } \frac{Y_\gamma}{\bar{Y}_\gamma} \right)\,
 ( \delta_{\mu\alpha}  - \delta_{\mu\beta} -\delta_{\mu\gamma})
\sum_{a\in\alpha, c\in\beta, d\in\gamma}  \bar{N}_{a \to c,d} \;
 \,, 
 \label{eq:decay}
\end{equation}
where
\begin{equation}
   \bar{N}_{a\to c,d} = \frac{Tg_a}{2\pi^2}m_a^2 \Gamma^0(a\to c,d)K_1\left(\frac{m_a}{T}\right) 
\end{equation}
and $\Gamma^0(a\to c,d)$ is the corresponding (zero-temperature, in-vacuum) partial width. 

Including decay processes in Eq.~\eqref{eq:dYdT} can lead to double-counting when the particle that decays also appears as an $s$-channel resonance in a $2\to 2$ process. In order to remedy this situation, in \micro\ we adopt the following solution: in sector $0$ we do not consider particle decays but, instead, carefully integrate the corresponding $s$-channel resonance; that is, we split the integration region into three sub-regions, thus ensuring that the pole (and, hence, the decay contribution) is properly accounted for. For DS particles, on the other hand, by default we include the decay terms in Eq.~\eqref{eq:dYdT} and exclude the region close to an $s$-channel resonance when integrating the collision term for $2\to 2$ processes, in order to avoid double counting. The region excluded from integration corresponds to 
\begin{equation}
      |s-m^2| < x\, m\, \Gamma \,,
\end{equation}
where $m$ is the particle mass, $\Gamma$ the particle's total  decay width and $x$ is the cut parameter, which in the code is named {\tt decayCutPar}. The default value for this parameter, $x\equiv$ {\tt decayCutPar = 5}, can be modified by the user. The processes taken into account in the width calculation include up to four particles in the final state. The user can also exclude the decay term in the evolution equation  by setting 
\begin{equation}
  {\tt ExcludedForNDM}={\tt ``DMdecay"} \,.
\end{equation}   
When this option is chosen, the region around the $s$-channel resonance is taken into account.
By default, when 2-body decays are present, the 3-body processes are not computed by \micro. However, there is a switch, which allows the user to include 3-body processes in all cases as explained in section~\ref{sec:functions}. 

\subsubsection{Including co-scattering}
\label{sec:evolutioneqsCDFO}

So far we have only considered WIMP-like DM candidates. In general, Eq.~\eqref{eq:dYdT} also contains co-scattering terms corresponding to scattering ($\chi\, \rm{SM} \to \chi'\, \rm{SM}$) or decay ($\chi' \to \chi\, \rm{SM}$) processes, both of which concern transitions of particles in one DS ($\chi$) into particles of another sector ($\chi'$) that also involve bath particles. Here $\chi$ and $\chi'$ may have the same transformation properties under the symmetry group, \textit{i.e.}, according to our previous definitions, be part of the same symmetry sector, yet they may belong to different DSs if they are not in thermal equilibrium. 

Our treatment of co-scattering effects begins with the observation that scattering processes $\mu, 0 \to \nu, 0 $ and decays $\mu \to \nu, 0 $ enter the evolution equations in a similar manner, namely
\begin{equation}
\label{eq:Ycoscat}
 3H\ent \frac{dY_{\mu}}{d \ent} \approx (Y_\mu -Y_\nu\frac{\bar{Y}_\mu}{\bar{Y}_\nu})   \,
\Gamma_{\mu\to\nu} 
\end{equation}
where we have defined
\begin{eqnarray}
\Gamma_{\mu\to\nu} =&&
      Y_0 \langle\sigma_{ \mu 0 \nu 0} v \rangle \ent(T) \\ \nonumber
& +&
   \frac{ \sum \limits_{a\in\mu, c\in\nu} g_am_a^2 \Gamma^0(a\to c,0)K_1\left(\frac{m_a}{T}\right)  + \sum \limits_{a\in\nu, c\in\mu}   g_a m_a^2 \Gamma^0(a\to c,0)  K_1\left(\frac{m_a}{T}\right) }   { \sum \limits_{a\in\mu} g_am_a^2 K_2\left(\frac{m_a}{T}\right) } \ \ .
   \label{eq:cowidth}
\end{eqnarray}
The quantity $\Gamma_{\mu\to\nu}$ can be seen as an \textit{effective decay rate} between sectors $\mu$ and $\nu$ and
\begin{eqnarray}
\bar{Y}_\mu \Gamma_{\mu \to \nu} &=& \bar{Y}_\nu \Gamma_{\nu \to \mu} \,, \\
\frac{ d (Y_\mu +Y_\nu)}{d\ent} &=&0 \,.
\end{eqnarray} 

The rate of the decay processes in Eq.~\eqref{eq:cowidth} slightly decreases at large temperatures because of the Lorentz factor $K_1(m_a/T)/K_2(m_a/T)$, while the rate of conversion processes  in general  increases linearly with temperature~\cite{Alguero:2022inz,Garny:2017rxs,DAgnolo:2017dbv}.\footnote{Since $H(T)$ increases as $T^2$,  co-scattering processes do not guarantee that chemical equilibrium is maintained at high temperatures.} 
Note that $Y_0$, which corresponds to the abundance of bath particles, always equals its equilibrium value,
$Y_0=\bar{n}_0/\ent(T)$. The dependence on $n_0$ in Eq.\eqref{eq:dYdT} is cancelled by $n_0$  in Eq.\eqref{vSigmaNExp}, hence, for simplicity, in {\tt micrOMEGAs} we set 
\begin{equation}
 \bar{n}_0=\ent(T),\;\;\; Y_0=1
\end{equation}
without loss of generality. 

For the computation of co-scattering, the user defines which particles belong to which sector, say $1$ and $2$. If all particles are  in thermal equilibrium but one of them is nevertheless assigned to sector 2, which contains no other particle, two abundance equations will be solved and should give the same result as the single abundance equation, that is $Y_{\rm final}{\rm(heavier\:particles)}=0$ and $Y_{\rm final}{\rm (lightest\:particle)}=Y_{\rm final}{\rm (single\:equation)}$ of the single equation. 

We should also point out the fact that, in general, in the co-scattering phase kinetic equilibrium may be lost; when this is the case one needs to solve the full momentum-dependent Boltzmann equation, see \textit{e.g.}~\cite{Brummer:2019inq} for specific examples. It is the responsibility of the user to verify that kinetic equilibrium is maintained for the considered parameter values, so that a treatment based on the momentum-integrated Boltzmann equations constitutes a good approximation.

\subsubsection{Including freeze-in}
\label{sec:evolutioneqsFI}

Lastly, the previous equations (Eqs.~\eqref{eq:dYdT}, \eqref{eq:decay}, \eqref{eq:Ycoscat}) can also be used for freeze-in production of DM, in which case  the initial conditions have to be set differently. In particular, for the freeze-in case we assume that the initial abundance vanishes,  
hence, the terms corresponding to DM production through pair annihilation of bath particles or through decays of heavier particles  are dominant. Nevertheless, the terms corresponding to DM annihilation are included, contrary to the assumption adopted when solving the one-component FIMP case. Note that here we assume a Maxwell-Boltzmann statistical distribution for all particles, and kinetic equilibrium is assumed throughout.

\subsection{Solution to the evolution equations and validation of the code}

Let us now describe in some more detail the way in which the Boltzmann equations describing the evolution of all relevant abundances in multi-component DM scenarios are solved in \microsix. In doing so, we will also introduce the main routines that can be used in order to solve these equations and compare their numerical results with well-tested, pre-existing ones in the appropriate regimes. More details concerning the utilisation of these functions will be given in Section \ref{sec:functions}.

First, the initial conditions which are necessary for the resolution of the evolution equations can either be set automatically or manually. The former choice, which is followed by the routine {\tt darkOmegaN}, provides correct results only if all DM particles are WIMPs. In order to find the starting temperature, we transform the differential equations for the abundances into linear equations with respect to 
\begin{equation}
  \Delta Y_i  =  \bar{Y}_i(T) - Y_i(T) \;,
\end{equation}     
ignoring derivatives and higher order terms in  $\Delta Y_i$. For each component, the initial integration temperature, \verb|Tstart|, is set by the condition
\begin{equation}
\label{eq:tstart}
  \Delta Y_i({\tt Tstart}) < 0.1 \times \bar{Y}_i({\tt Tstart}) \,.
\end{equation}
In practice, to find the initial temperature {\tt darkOmegaN} starts at a low temperature and increases the temperature until Eq.~\eqref{eq:tstart} is satisfied. If this equation is not satisfied even for temperatures of the order of the DM mass, the calculation stops and an error code is written. This will occur, for example, if one of the DM components is a FIMP. In this case, both the initial temperature and the initial abundances have to be defined explicitly and manually. This can be done through the routine {\tt darkOmegagaNTR}, which allows the user to set the initial conditions for the abundances at a given initial temperature. Therefore, this routine  should be used if at least one of the components is not in thermal equilibrium with the SM bath. Typically, one sets  $Y[i]=0$ for the FIMP components and $Y[i]= \bar{Y}[i]$ for WIMP components. The initial (``reheating'') temperature should preferably be set to a value which is not too high, since too large an initial temperature can lead to problems with the numerical resolution of \textit{stiff} equations.\footnote{In particular, in the case of WIMPs at high temperatures $Y_i$  is very close to $\bar{Y}_i$ and the standard  Runge-Kutta methods lead to small step oscillation of $Y_i$ around $\bar{Y}_i$.} In \micro\ we use the Rosenbrock algorithm~\cite{PressNumerical} for the resolution of stiff equations. This method finds a solution for points where the standard Runge-Kutta method fails. In the current version, we use the C code presented in \cite{PressNumerical}.

The solution of the evolution equations follows the procedure described in~\cite{Belanger:2014vza}. The equations are solved in an interval \verb|[Tend,Tstart]| where \verb|Tend| can be defined by the user. The default value is \verb|Tend|$=10^{-3}$ GeV since the evolution of the abundance typically stops at higher temperatures. 
However, decays of long-lived DS particles can take place at lower temperatures, to properly take these into account it is preferable to choose a smaller value for {\tt Tend}.
The total relic density reads\footnote{In this Section we depart from our usual convention of labelling different DSs with Greek indices.}
\begin{equation}
\Omega h^2= \sum_i \Omega_i h^2= \sum_i 2.742\times 10^8 Y_i M(\chi_i) \,,
\end{equation}
where $M(\chi_i)$ is the mass of the DM in the $i^{th}$ sector. In the code it corresponds to the array element {\tt McdmN[i]}, i=1,...{\tt Ncdm}, where {\tt Ncdm} is the total number of DSs. We define
 $\xi_i$ as  the relative contribution of DM$_i$ to  the total DM density, $\xi_i=\Omega_i/\Omega_{tot}$, which in the code corresponds to the array element {\tt fracCDM[i]}. 
 
When solving the $N$-component equations, we assume that all particles which are in thermal equilibrium follow Maxwell-Boltzmann distributions. In Ref.~\cite{Belanger:2018ccd} it was shown that taking into account Fermi-Dirac or Bose-Einstein statistical effects can lead to non-negligible corrections to the relic density of FIMPs -- up to nearly a factor two when freeze-in is dominated by annihilation processes, while the correction is at the level of a few percent when DM production is dominated by decay processes. Dedicated routines for the freeze-in production of single-component DM are also available in \micro, as presented in~\cite{Belanger:2018ccd}, through which statistical effects can be fully accounted for. We recommend to use these for a more precise evaluation of the FIMP component. A discussion of some specific issues related to freeze-in can be found in Section~\ref{sec:FI}.
\\
\\
Extensive tests of all $N$-component functions were performed within the framework of different models: the singlet scalar model, the inert doublet plus singlet model with a $Z_4$ symmetry (Z4IDSM)~\cite{Belanger:2021lwd} as well as the two singlet model with a $Z_5$ symmetry (Z5M)~\cite{Belanger:2020hyh}. The former was used in order to ensure that, when appropriate, the results of {\tt darkOmegagaNTR} match the predictions of existing \micro\ routines in the simplest case of one-component DM freeze-in, and excellent agreement was found. The latter two models, on the other hand, were used in order to test both the case of multiple WIMPs and WIMP+FIMP DM scenarios.

\begin{figure}[htb]
\centering
\includegraphics[scale=0.48]{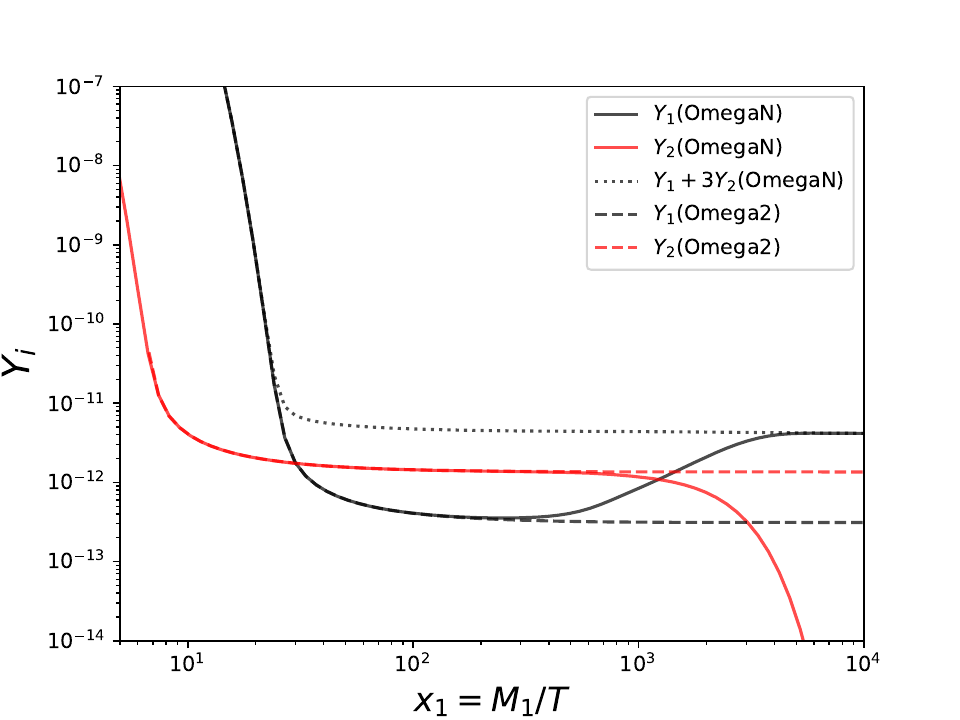}
\includegraphics[scale=0.48]{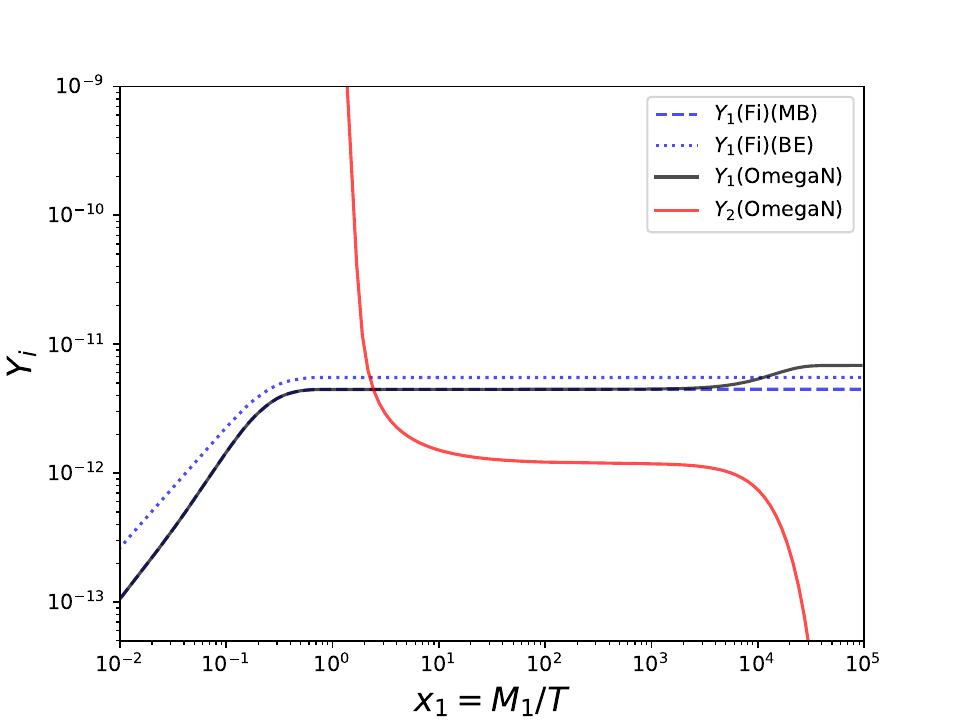}
\caption{ Left: Comparison of the abundances ($Y_i$) as computed with {\tt darkOmega2} (dashed) and {\tt darkOmegaN} (full) in the Z5M  for the parameters  specified in Eq.~\eqref{data:figom2}.  Also shown is $Y_1+3Y_2$ as computed with {\tt darkOmegaN}  without decays (dotted).
 Right: Comparison of the abundance $Y_1$  as computed with {\tt darkOmegaNTR} (full, black) and {\tt darkOmegaFi}  assuming Maxwell-Boltzmann (dashed) or  Bose-Einstein quantum statistics (dotted) in the Z5M for the parameters specified in Eq.~\eqref{data:figomfi}. Also shown is $Y_2$ (red).}
\label{fig:om2omN}
\end{figure}

Concretely, focusing on the Z5M with two WIMP-like DM candidates, good agreement was found between the result of {\tt darkOmega2}, a \micro\ routine dedicated to two-component WIMP-like DM scenarios \cite{Belanger:2014vza}, and {\tt darkOmegaN} for all cases in which the decay term is not relevant (the decay term is included only in the latter). In order to illustrate this, let us we consider the case in which the relevant couplings are such that $\phi_1$ and $\phi_2$ are in thermal equilibrium with the SM bath and there is a small coupling $\lambda_{31} \phi_1^3 \phi_2$ which leads to the decay of $\phi_2\rightarrow \phi_1\phi_1\phi_1$ after the freeze-out of $\phi_1$. The parameters of the model~\cite{Belanger:2020hyh} are chosen as
\begin{eqnarray}
&&M_1=100~{\rm GeV} \,,\; M_2=350~{\rm GeV} \,,\; \lambda_{S1}=0.2\,,\;\lambda_{S2}=0.14\,,\; \nonumber\\
 && \lambda_{31}=10^{-9},\; \lambda_{41}=\lambda_{42}=0.001\,,\; \mu_{SS1}=\mu_{SS2}=\lambda_{31}=\lambda_{412}=0 \,.
\label{data:figom2}
\end{eqnarray}
In the left-hand side panel of Fig.~\ref{fig:om2omN} the abundances computed with {\tt darkOmega2} (dashed) are compared against those computed with {\tt darkOmegaN} (solid).  We find that $Y_1$ and $Y_2$ computed with either routine are in perfect agreement down to the temperature at which the decay takes place. At this point $\phi_2$ disappears, an effect which cannot be captured by {\tt darkOmega2}, and the abundance of $\phi_1$ (dotted) is given by $Y_1 ({\rm with~decay})=Y_1 ({\rm w/o~decay})+3Y_2({\rm w/o~decay})$, where the low-temperature abundances are computed by {\tt darkOmegaN} including or not the decay term. Similar agreement was found within the framework of the Z4IDSM model.
 
Passing to the case of mixed WIMP/FIMP DM, we first considered the case of such a setup in the Z4IDSM model. Ignoring statistical distribution effects, good agreement was found between the results of {\tt darkOmegaNTR} and {\tt darkOmegaFI}, the \micro\ routine dedicated to the computation of the DM abundance in single-component freeze-in scenarios, for the relic density of the FIMP. These tests were also repeated in the Z5M for the case in which $\phi_1$ is a FIMP while $\phi_2$ remains weakly interacting. Our results are presented in Fig.~\ref{fig:om2omN} (right panel), where we compare different computations of the abundances for the same parameter choices as in Eq.~\eqref{data:figom2} except for
\begin{equation}
 \lambda_{S1}=10^{-11},\;  \lambda_{S2}=0.15\,,\; \lambda_{31}=0\,,\; \mu_{SS1}=10^{-9} \,.
\label{data:figomfi}
\end{equation}
The last coupling corresponds to  a term $\mu_{SS1} \phi_2 \phi_1\phi_1^\dagger$ and induces the decay of $\phi_2$, hence the FIMP remains the only DM candidate. Here the abundances of $\phi_1$ and $\phi_2$ computed with {\tt darkOmegaNTR} (black and red solid lines, respectively) are shown, and $Y_1$ can be compared with the one computed with {\tt darkOmegaFi} assuming a Maxwell-Boltzmann distribution for all bath particles (blue-dashed curve). We find that, as expected, the two results are in perfect agreement down to the temperature where the decay of $\phi_2$ takes place, around $x=10^4$. Recall that the FI routine only considers FIMP production from particles in thermal equilibrium with the SM, and as such it does not include the decay of the WIMP after it freezes-out (in other words, {\tt darkOmegaFi} continues to assume an exponential suppression of $\phi_2$ even below its freeze-out temperature), hence a smaller value of $Y_1$ is found. This figure also shows that, once Bose-Einstein statistics are properly accounted for in the freeze-in calculation (blue-dotted curve), $Y_1$(BE) exceeds $Y_1$(MB) by about 30\% as expected~\cite{Belanger:2018ccd}. Lastly,  we also found that the value of $\Omega_2$ computed with the standard single-component \micro\ routine {\tt darkOmega} while assigning $\phi_1$ to the list of feeble particles coincides with the value computed with {\tt darkOmegaNTR} when ignoring decays.

\section{The role of decay and co-scattering processes in maintaining  chemical equilibrium.} 
\label{sec:coscat}

One of the underlying assumptions that we made when writing the Boltzmann equations is that chemical equilibrium is maintained within each DS. 
In the presence of small couplings this might not be the case, in particular in scenarios  where co-scattering is responsible for DM formation~\cite{Alguero:2022inz}.
In these scenarios, there are two types of processes which are
relevant for chemical equilibrium between dark particles $\chi,\chi'$: decays ($\chi' \to \chi\,\SM$) and inelastic scattering  ($\chi\,\SM \to \chi'\,\SM'$).  
In general the decay processes are responsible for chemical equilibrium at low
temperatures, while inelastic scattering processes  maintain chemical equilibrium  at high temperatures. 
In some cases, a small mass gap between $\chi'$ and $\chi$ leads to a small decay width that is not sufficient to maintain chemical equilibrium. However
for a  DS containing many particles, it is also possible that chemical equilibrium is restored  through interactions with a third particle $\chi''$ with a large mass gap with the DM $\chi$. Then, interactions between $\chi \leftrightarrow \chi'' \leftrightarrow \chi'$ can lead to chemical equilibrium between $\chi$ and $\chi'$. 

Below we propose a method to  determine  
whether or not  decay and co-scattering terms are strong enough to guarantee chemical equilibrium within  one DS.
For this  purpose we   divide  a given sector  into two sets, say $A$ and $B$, 
and we assume that   for set  $A$  the abundance equation contains 
only decay and co-scattering terms,  Eq.~\eqref{eq:Ycoscat}.
\\
\\
We introduce a parameter $\Delta$ to characterize the deviation from equilibrium 
\begin{equation}
   Y_A= Y_B\frac{\overline{Y}_A}{\overline{Y}_B}(1+\Delta) \,.
\end{equation}
Substituting $Y_A$ in Eqs.~\eqref{eq:dsdT} and \eqref{eq:Ycoscat}, we can  estimate $\Delta$ as 
\begin{equation}
   \label{Delta}
  \Delta=\frac{H}{\Gamma_{A\to B}} \frac{d  \left (Y_B \overline{Y}_A/ \overline{Y}_B\right)} { d\log{T}}    \,.
\end{equation}
Thermal equilibrium means that $\Delta \ll 1$. When   $T$ is larger than the freeze-out temperature and assuming that particles in 
subset $B$ are in thermal equilibrium  with the bath, then $Y_B\approx  \overline{Y}_B $ and
\begin{equation}
     \Delta=\frac{H}{\Gamma_{A\to B}} \frac{d\log( \overline{Y}_A)} { d\log{T}} \approx \frac{H}{\Gamma_{{A\to B}}}
\frac{M_A} {T}   \approx \frac{H}{\Gamma_{{A\to B}}} X_f  \,.
\end{equation}
Here $M_A$ is the mass of the lightest particle  in subset $A$. 
Thus, the condition for chemical equilibrium is $\Gamma_{{A\to B}}/H \gg X_f$.\footnote{
$H(T)\approx 10^{-18}T^2/1$~GeV and $X_f\approx 20$ for $M_{\rm cdm}\approx 100$~GeV, thus $\Gamma=10^{-14}$~GeV is  sufficient to maintain chemical equilibrium.} 
Taking all possible ways to split a given sector the minimal value for $\Gamma_{{A\to B}}/H$ is computed in the {\tt checkTE} routine;  if this condition is satisfied, then all particles in the sector are in chemical equilibrium. If not, the DSs have to be further split in subsectors before computing the relic density.
\\
\\
After freeze-out
$Y_B$ is approximately constant and Eq.~\eqref{Delta} reads 
\begin{equation}
    \Delta=\frac{H}{\Gamma_{A\to B}} \frac{M_A-M_B}{T}  \,.
\end{equation}
Note that co-annihilation  does not play a role in  DM formation when $(M_A-M_B)/T \gg 1$ even if $A$ and $B$ are not in chemical equilibrium. The  condition   
$\Gamma_{A\to B}/{H} \gg 1$ 
is therefore sufficient   for small temperatures.  

\section{Some issues with singularities}
\label{sec:singularities}

The formalism presented in the previous Sections is fairly general and can be applied to a broad spectrum of extensions of the Standard Model of particle physics. However, in concrete realizations of (conversion-driven) freeze-out or freeze-in, \textit{i.e.}\  within a specific DM model, there can be complications which arise due to the existence of singularities which appear, in particular, when computing processes that involve $t$-channel exchange. In this Section we point out a few such cases and clarify the way they are treated in \micro.

\subsection{$t$-channel vector exchange and freeze-in}
\label{sec:FI}

The first type of singularity that can appear in practical computations is a $t$-channel, $p_T \to 0$, singularity in scattering processes in the limiting case where the mass $M$ of the exchanged particle is vanishingly small compared the the total energy $\sqrt{s}$ of the process: $\sqrt{s}\gg M$. As discussed in the Appendix of the {\tt micrOMEGAs5.0}\cite{Belanger:2018ccd} this is particularly acute for the exchange of a spin-$1$ and to a lesser extent for a spin-$1/2$ particle. In the former, the {\it total integrated cross section} at asymptotically high energies scales to a constant  $\propto 1/M^2$ (power singularity) and does not decrease with energy. In the context of DM freeze-in, for instance, this  leads to a DM abundance which exhibits a linear dependence on the reheating temperature even though the model is renormalizable. For spin-$1/2$ exchange the scaling is $\propto  \log(s/M^2)/s$ (logarithmic singularity), but nonetheless decreasing with energy. 

However, in an actual cosmological context, this behaviour is very much tamed by taking into consideration temperature effects which are usually small. When the $t$-channel exchanged particle is part of the thermal bath it acquires an effective temperature-dependent mass, $M_{\eff}(T)^2=M^2+ F_M (T)$, where $F_M$ is the temperature functional which is model-dependent~\cite{Weldon:1982aq,Rychkov:2007uq}. $M_{\eff}(T)$  increases with temperature such that the total cross section at high energies scales as $1/M_{\eff}^2(T)$ and therefore
\textit{decreases} with temperature. In order words, there is a typical cutoff of the order of this thermal mass (or of the temperature). To take this effect into account, so that the cross sections mediated by bath species have the correct temperature dependence, one might be tempted to just replace {\em by hand} the {\em default zero-temperature} mass $M$ (\textit{i.e.}\ the one implemented in the \micro\ model files) by its $T$-dependent expression $M_{\eff}(T)$. This can be a particularly hazardous solution, especially in the case of {\it gauge} spin-$1$ mediator exchange which is the most often encountered manifestation for the occurence of this singularity. Making such a blind replacement will devoid the theory of its gauge invariance leading to a worse problem: a loss of unitarity with cross sections not being constant at asymptotic energies but even increasing with energy. A case in point in the SM is $\gamma \gamma \to W^+ W^-$. While this problem could, in principle, be avoided on a process-by-process basis, it is highly impractical for a tree-level-based multipurpose code in which many processes are generated for the same point in parameter space in order to compute the DM relic density. For this reason, in \micro\ we instead simulate the asymptotic 
$1/M_{\eff}^2(T)$ by applying a temperature-dependent $p_T$ cut on the total cross section (which is  generated with the original model file defined at zero-temperature with mass $M$). Gauge invariance is not broken as all $T=0$ tree-level contributions to any process are {\em automatically} generated and without having to single-out any particular $t$-channel Feynman diagram. We will show below how the $p_T$ cut is chosen in order for the {\em full} asymptotic behaviour to be recovered and that a $p_T$ temperature-dependent cut constitutes a very good approximation.

In order to make the connection between an effective $p_T$ integration cut and the thermal mass, let us take as an example a Standard Model process involving $t$-channel \mbox{spin-1} exchange: $\epem \to \nu_e \bar {\nu_e}$. In this example, it is actually sufficient to only consider the gauge invariant $t$-channel contribution.\footnote{There is also an $s$-channel contribution to this process but this is the same as the one for $\nu_{\mu,\tau}$ production, which does not involve $t$-channel diagrams and is therefore trivially gauge independent with massless neutrinos.} In this case, the only relevant mass is that of the $W$ boson, $M_W$. The differential cross section reads 
 \beqn
 \label{eq:spin1Wenue}
 \frac{{\rm d}\sigma}{{\rm d} \Omega} = \frac{\alpha^2}{16 s_W^4} \frac{1}{s}\;\; \frac{\cos^4 (\theta/2)}{\left(\sin^2 (\theta/2)+M_W^2/s \right)^2}, \;\;\;\; 
 {\rm d} \Omega=2\pi {\rm d} (\cos \theta) \,,
 \eeqn
where $\theta$ is the scattering angle, $s_W$ is the sine of the weak angle, $\alpha$ is the fine-structure constant.

The integrated cross section can be obtained analytically even in the presence of a cut, say $c$, on the (cosine of the) scattering angle 
\begin{align}
\label{xs_full}
\sigma(\sqrt{s},M_W,c)&= \hat{\sigma}_{e \nu_e} \; \int_{-1+c}^{1-c} \; \text{d}x \frac{(1+x)^2}{(1-x+2 \mu^2)^2}, \quad \mu^2=M_W^2/s, \quad  \hat{\sigma}_{e \nu_e}=\pi \alpha^2/ (8 s_W^4 s),
\nonumber \\
&=\hat{\sigma}_{e \nu_e} \left( 4- 4 c +\frac{c^2}{-2+c-2\mu^2} + \frac{(c-2)^2}{c+2 \mu^2}+4 (1+\mu^2) \log \bigg( \frac{2\mu^2+c}{2\mu^2+2-c} \bigg) \right) \nonumber \\
&= 4 \hat{\sigma}_{e \nu_e} \left( \frac{1}{2 \mu^2+c} + \log \bigg( \frac{2 \mu^2+c}{2} \bigg)  + \mu^2 \log \bigg( \frac{2 \mu^2+c}{2} \bigg) + 1 - c \right.   \\ 
  & \left. \qquad + \frac{c^2}{4(-2+c-2\mu^2)} + \frac{c (c-4)}{4 (c+2 \mu^2)}- (1+\mu^2) \log \bigg( 1+\mu^2-c/2 \bigg) \right) \, . \nonumber 
\end{align}
We have written the above expression as the sum of the leading (universal, spin-1) power singularity ($1/(2 \mu^2+c)$), a subleading logarithmic singularity (``spin-1/2'', $\log (2 \mu^2+c)$) and the remaining process-dependent constant term in which  we can take the limit $\mu \to 0$ and $c \to 0$ at will. In that limit the last line is  $+1$.  The crucial observation here is that the regulators $\lbrace \mu, c \rbrace$ only enter the leading and subleading singularities through the same combination $2 \mu^2+c$. This demonstrates our main point, we can make the identification 
\beqn
 2 \mu^2+c=2 \mu^2+c(T) \equiv 2 \mu^2_{{\rm eff}}=2 \mu^2(T)
\eeqn
An effective temperature-dependent mass  $\mu^2_{{\rm eff}}(T) $ for the asymptotic behaviour of the cross section calculated with no-cut ($c=0$) is equivalent to a zero-temperature mass calculation but with a temperature dependent cut, $c(T)$. 
\\
\\
Of course, for this procedure to make sense (\textit{i.e.}, with a cut), 
\beqn
0 < c <1 \, , \quad c \equiv 2 (\mu_{{\rm eff}}^2- \mu^2)
\eeqn
must hold. Note that the argument applies equally for a spin-1/2 exchange. The asymptotic, high-energy, constant cross section behaviour without cut ($c=0$, $s \gg M_W^2$) is recovered as 
\beqn
\label{asympt_enue}
\sigma_{e \nu_e}^{\rm asymp}=\frac{\pi \alpha^2}{4 s_W^4} \frac{1}{M_W^2}= 2 \hat{\sigma}_{e \nu_e} \bigg(\frac{s}{M_W^2}\bigg) \,.
\eeqn
Now, we can also trade the cut on the integration angle, $c$,  in Eq.~\eqref{xs_full} with a cut on the transverse momentum, $p_T$, through
\beqn
\sin^2 \theta= \frac{4 p_T ^2}{s} \,. 
\eeqn
For small enough $c$, 
\beqn
c=\frac{2 p_T^2}{s} \,.
\eeqn
Keeping only the leading terms  in $1/\mu^2, 1/c (1/p_T^2)$, we obtain
\beqn
\label{xs_full_approx}
\sigma(\sqrt{s},M_W,p_T)& \simeq &4 \; \sigma_{e \nu_e} \; \Bigg( \frac{1}{2} \frac{s}{M_W^2+p_T^2}+ 1+\log \frac{M_W^2+p_T^2}{s} \Bigg) \nonumber \\
& \simeq & \frac{\pi \alpha^2}{2 s_W^4} \Bigg( \frac{1}{2 M_W^2(T) } + \frac{1}{s} \bigg( 1 +  \log \frac{M_W^2(T)}{s} \bigg) \Bigg) \,.
\eeqn 
All in all, from the previous example we see that the leading effects of a (finite-) temperature-dependent mass can be encapsulated by the imposition of a $p_T$ integration cut on the zero-temperature computation. Then, the $p_T$ cut that corresponds to the correct $T$-dependent mass  sourced from Debye screening is 
\beqn
p_T^2= M_W^2(T) -M_W^2(T=0) \,.
\eeqn
We have thoroughly verified 
 that the identification we have made works also in the case of more involved spin-1 exchange processes in the SM and tested a few examples with spin-1/2 exchange. 
\\
\\
The temperature-dependent correction, $F_M(T)$,  for the mass can generally be cast as 
\beqn
  M_{\text{eff}}(T)^2=M^2 + \kappa^2 T^2 \,.
  \label{eq:meff}
\eeqn 
With all SM particles in the bath, the Debye correction for the $W$ mass approximates as~\cite{Weldon:1982aq,Rychkov:2007uq}
 \beqn
 M_W^2(T)= M_W^2+\kappa^2 T^2 \quad \text{with} \quad \kappa^2=\frac{11 \pi}{3 s_W^2} \alpha \,.
 \eeqn
 
 General formulae for $\kappa^2$ depend on the model of DM and the nature of the exchanged particle. We leave it to the users to define their own model by providing a table for the $\kappa^2$s. In \microsix\ the possibility is given to appropriately fill  the array {\tt Tkappa[k]}~$\equiv \kappa$   where {\tt k} corresponds to the $k$-th particle in the model and {\tt Tkappa[k]} is identified with $\kappa$ 
 in the temperature-dependent expression of its squared mass in Eq.~\ref{eq:meff}. When multiple $t$-channel propagators appear in a process, the strongest cut is kept. More details about the array {\tt Tkappa[k]} will be given in Section \ref{sec:functions}.

\subsection{$t$-channel poles and thermally-induced widths}

Another type of singularity can appear when computing matrix elements for $2\to2$ processes which receive contributions from $t$-channel diagrams in which the exchanged particle is stable (\textit{i.e.}, its in-vacuum, zero-temperature decay width is zero). A relevant example that was already encountered in the very early version of \micro\ \cite{Belanger:2004yn}  is the supersymmetric process $\tilde{\mu} \chi \to \mu h$ (with an incoming unstable $\tilde{\mu}$) through $t$-channel exchange of neutralino DM ($\chi$). For low-energy collisions the corresponding $t$-channel pole is found in the case where $\chi$ from the decay $\tilde{\mu}\to \mu\chi$ collides with another $\chi$ to produce a Higgs boson ($h$). Note that there is double - counting between this process and the one where a pair of neutralinos annihilate through exchange of a Higgs boson.  The pole in the physical region leads to a divergence when integrating over the scattering angle. 
In the collider context, the meaning of such divergences was discussed in \cite{Melnikov:1996iu}. In a plasma, on the other hand, we expect the \textit{thermal width} of the $t$-channel particle (which does not vanish even for a stable particle) to regularise the integral.\footnote{Physically, such a thermal width would reflect the fact that in a plasma, even a stable particle has a finite absorption probability. See for instance \cite{Boyanovsky:2004dj,Grzadkowski:2021kgi}.}
In earlier versions of \micro, this issue was addressed by introducing a small width, $\Gamma=M/100$, for $t$-channel particles. This technique is still used for the functions {\tt darkOmega} and {\tt darkOmega2} and we now automatically extend it to  {\tt darkOmegaN}  when the code finds a  $t$-channel  pole in the vicinity of the physical region.

\subsection{Infrared problem in co-scattering processes.} 

If the decay $\chi_2\to \chi_1 X$ (where $X$ is a bath particle) is kinematically open and one of the  DS particles, $\chi_2$  or $\chi_1$, has an electric and/or colour charge, then the cross section of the co-scattering process  $\chi_2, \gamma/g \to \chi_1, X$  in the soft photon or gluon region can be approximated as 
\begin{equation}
\label{infra22}
   \sigma \approx \alpha\frac{\Gamma_{\chi_2\to \chi_1,X}}{E^3} \,,
\end{equation}
where $E_{\gamma/g}$ is the photon/gluon energy and $\alpha$ stands for the electromagnetic or strong coupling constant. This tree-level contribution is infra-red divergent and shows as a logarithmic divergence upon convolution in Eq.\eqref{nEvents22}. In~\cite{Alguero:2022inz} it was proposed to apply a cut on the centre-of-mass energy of 
($g,\chi_2$), 
effectively removing the soft region and hence  the infra-red divergence but at the expense of having a result which is infra-red cut-dependent. Although the dependence on the ad-hoc cut-off is only logarithmic we deem this not to be satisfactory. 

It has been known for a long time that, at $T=0$,  the one loop correction where the photon is not external but internal in the loop has also an infra-red divergence that cancels the real soft photon induced process. In fact, the two processes define an infra-red finite corrected width. 
This cancellation also occurs at $T\neq 0$ when the process takes place in a thermal bath. 
The details of the cancellation have been first worked out in \cite{Czarnecki:2011mr}.

The calculation of thermal corrections to the decay width  both of charged particles and of neutral particles decaying into two charged ones was performed in  \cite{Czarnecki:2011mr, Beneke:2016ghp, Ghiglieri:2019lzz} with an expansion in $T/M_{\chi_2}$. For DM freeze-out $T/M=1/x_f\approx 1/20$ is a small parameter and, therefore, the results of those papers can be exploited. For all cases considered, the thermal corrections to the decay width were found to be  small and negative. We can therefore safely ignore such corrections and at the same time not consider such massless gauge boson-initiated co-scattering.  Therefore, in \micro\ we simply exclude such processes. 

\section{Other improvements}
\label{sec:other}

\subsection{3-body and 4-body final states}

When the cross section for DM annihilation into two-body final states is kinematically suppressed, the contribution from 3-body or even 4-body final states can be important. A well-known case is DM annihilation into SM gauge boson pairs just below threshold. In previous \micro\  versions~\cite{Belanger:2013oya}, a switch ({\tt VWdecay,VZdecay}) could be turned on to include the annihilation into one or two virtual gauge bosons in the relic density calculation. 

In \microsix, this is generalised to other heavy states other than $W/Z$, for example those involving top quarks or new non-Standard Model particles. Concretely, the functions \verb|vSigmaPlus23| and \verb|vSigmaPlus24| can be used to calculate the  off-shell contribution to  specific $2\to 3$ and $2\to 4$ processes defined by the user (see the description in Section~\ref{sec:functions}). This contribution is then added to the  DM annihilation cross section that includes all possible 2-body final states and is used for the relic density calculation in single-component DM models.

 To be able  to include such multi-body final states while maintaining a fast enough code, we make some approximations. We  do not compute the full 4-body final state but only the contribution from the diagrams with one off-shell particle. For example, consider the process $\chi\chi \to B B'$, where $B$ and $B'$ stand for any sector-0 particle in the model. \verb|CalCHEP| is used to compute $\chi \chi \rightarrow B f f'$, where $B$ corresponds to the particle with the smallest width and $f f'$ corresponds to  the main decay channel of $B'$.\footnote{Here, we set aside the issue of possible gauge invariance breaking  which may arise, for example, with virtual $W$ and $Z$ final states. The way through which this problem can be circumvented in \micro\ was  described in \cite{Belanger:2013oya}.} 
Far below threshold, when both particles are virtual,  the total cross section should be the sum of $\chi \chi \rightarrow B' f_1 \bar{f_1'}$ and $\chi \chi \rightarrow B f f'$, each rescaled 
by the corresponding branching fractions, while very near threshold  either of the 3-body result should be close
to the 2-body result.  A more robust result would, therefore, require to compute the 4-body final state. Rather than doing this we use the recipe described in~\cite{Belanger:2013oya} to compute a $K$-factor to rescale the 3-body process.

 The user also has the possibility to compute the 4-body process associated with two off-shell particles.
 For this we choose  for each off-shell particle the decay channel with the largest branching fraction  (amongst the ones which involve only sector-0 particles) and again use \verb|CalcHEP|~\cite{Belyaev:2012qa} to compute the cross section for the process $\chi\chi' \rightarrow B^*(\to f_1 f_1')  B'^* (\to  f_2 f_2')$. 
The result is then divided by the  product of the respective branching ratios ${\rm Br}(B\to f_1f_1') \times {\rm Br}(B'\to f_2 f_2')$. 
 The computation of 4-body final states is CPU-time consuming and not well-adapted to large scans of parameter spaces, whereas the contribution from processes with two off-shell particles is typically not significant.

We have compared the results for the relic density obtained with and without the 3-body final states in two different scenarios, namely in the Inert Doublet Model (IDM)~\cite{Belanger:2013oya} and in the $U_1$ leptoquark model studied in~\cite{Belanger:2022kvj}.  We found a large decrease of the relic density close to threshold, as illustrated in Fig.~3 of Ref.~\cite{Belanger:2013oya} for the IDM model.
Moreover, we found that the relic density computed using only the $3$-body final states differs only by a few percent from the one including also the 4-body contribution, while the computation is at least a factor of 3 faster in terms of CPU time. 
Finally,  including the multi-body final states slightly above threshold (producing $B$, $B'$ on-shell) leads to a shift in the value of the relic density.
This shift varies from a few percent for a narrow resonance ($\Gamma/M=1\%$)  to 15\% for a wide resonance $\Gamma/M\approx 10\%$).

\subsection{Sommerfeld enhancement}
\label{subsec:Sommerfeld}
DM annihilation cross sections can be enhanced at low velocities through the so-called Sommerfeld effect~\cite{Hisano:2004ds}. This non-relativistic effect occurs when two heavy incoming particles interact with each other by exchanging a very light boson before pair-annihilation. In the case of particles interacting through a Yukawa potential, this Sommerfeld enhancement was calculated in, \textit{e.g.}, \cite{Iengo:2009ni} for the case of scalar and vector interactions. In \micro\ we follow the same prescription. We should warn the reader that this implementation of the Sommerfeld effect does not apply, in the current version, to the case where the initial system belongs to a multiplet, where co-annihilation takes place. 
The Lagrangians that describe the interactions of a Dirac particle ($f_D$)  or of a  charged  scalar field  ($\phi$)   with a scalar ($h$) or a vector ($v_\mu$) mediator, read
\begin{eqnarray}
 {\cal L}&=& e h \bar{f}_D f_D\,, \;\;\;\;  {\cal L}=e v_\mu \bar{f_D}\gamma^\mu f_D  \,, \nonumber\\
 {\cal L}&=&2e M_\phi h \phi\phi^*, \;\;\;\;  {\cal L}= i e v_\mu  (\partial^\mu \phi h \phi^* - \phi \partial^\mu \phi^*) \,.
 \label{eq:Lagrangian}
\end{eqnarray}
For Majorana particles, ${\cal L}$  contains an extra factor $\frac{1}{2}$.  The Sommerfeld enhancement factor  depends on the quantities
\begin{eqnarray}
   a &=& \alpha/v \,, \\
   b &=& \mu/m_r v  \,,  
   \label{eq:Sommerfeld}
\end{eqnarray} 
where $v$ is the relative velocity of colliding particles, $\mu$ is the mass of the mediator, $m_r=m_1 m_2/(m_1+m_2)$ is  the reduced mass of colliding particles of masses $m_1$ and $m_2$ and $\alpha={e^2}/(4\pi)$, with the coupling $e$ defined in the Lagrangian in Eq.~\eqref{eq:Lagrangian}. The routine that calculates the enhancement factor (\verb|Sommerfeld(a,b)|), see section~\ref{sec:functions}, is based on Eqs.(5.1)--(5.7) of \cite{Iengo:2009ni}.\\

The Sommerfeld factor for  pseudo-scalar  and axial-vector interactions is negligible~\cite{Agrawal:2020lea} and is not included in \micro.

\section{Improved limits on DM}
\label{sec:limits}
Apart from the upgrades in the DM relic density calculation described in the previous sections, in \microsix\ we also introduce several improvements in the computation of the experimental limits  from direct, indirect detection and cosmological observables other than the relic density. In this section we briefly describe these improved features.

\subsection{Limits from direct detection}

In multi-component DM models, it is straightforward to compare the direct detection cross section with the 90\% C.L. provided by experimental collaborations when only one component $\chi_i$ dominates the signal. In this case, one has to simply compare $\sigma^{\rm SI}_{\chi_i p} \xi_i$  or $\sigma^{\rm SD}_{\chi_i p/n} \xi_i$ with the limits provided by the experimental collaborations for a given DM mass. 

The situation becomes more involved in general multi-component DM scenarios and/or for different velocity distributions. In Ref.~\cite{Belanger:2020gnr}, some of us performed a recasting of four experiments, namely XENON-1T, PICO-60, CRESST-III and DarkSide-50.  The procedure described in this paper leads to an 
upper limit on the elastic scattering cross section which, generally, falls  within 10\% of the experimental limit in the case of one-component DM models. Similar precision is expected for 
 multi-component DM scenarios. The function \verb|DD_pval| described in Section~\ref{sec:functions} determines whether a given set of input parameters for a multi-component model exceeds or not the experimental limit.

\subsection{Photon spectra for Indirect detection of light DM}
\label{sec:ID}

DM can produce $\gamma$-rays by annihilating into other particles that either subsequently produce photons through final state radiation (FSR) / Inverse Compton Scattering, or which decay into photons.\footnote{Another possibility is DM annihilating into $X\gamma$, where $X$ is a neutral SM particle like $\gamma,\,Z,\,h$.}
In previous versions of \micro, the tables specifying the photon spectrum ($dN/dE$) for any pair of SM final states were obtained with {\tt Pythia\,6} and covered the mass range from 2~GeV to 5~TeV~\cite{Belanger:2010gh}. In this version, we also include the photon spectra for DM masses between 110~MeV and 2~GeV. These were obtained for the leptonic and hadronic channels using analytical formulae. For DM mass below 2 GeV, the possible hadronic final states produced by DM annihilation are dominated by pion and kaon pairs. The possible annihilation channels are listed below, along with the references that were used in order to obtain the relevant spectra:

\subsubsection{Light leptons} 

\begin{enumerate}
    \item {\bf Electrons: } DM annihilation into a pair of electrons only produces photons through FSR~\cite{Coogan:2019qpu}.
    \item {\bf Muons: } DM annihilation into a pair of muons can generate photons in two ways, (i) through FSR and (ii) through the radiative decay of muons~\cite{Coogan:2019qpu}. For sub-GeV DM, above the DM DM$\to \mu^+\mu^-$ kinematic threshold, the radiative muon decay dominates the photon spectra over FSR and the latter takes over as the DM mass increases.
\end{enumerate}

We have updated the relevant tables included in previous versions of \micro\ to include the possibility of DM pairs annihilating into $e^+e^-$ or $\mu^+\mu^-$ in order to extend the coverage in DM mass. Moreover, the radiative decays of the muon are included and we have improved the spectra from the electron channel at low energies. 

\subsubsection{Pions}

\begin{itemize}
    \item {\boldmath $\pi^0\pi^0 :$} The photon spectra for $\chi\bar{\chi}\to\pi^0\pi^0$ is generated through $\pi^0$ decays into a pair of photons, with a branching fraction $\sim 99\%$. In the rest frame of $\pi^0$, the photon spectrum is just a monochromatic line around $m_{\pi^0}/2$ while with  the appropriate boost in the galactic frame, it becomes a box-shaped spectrum~\cite{Boddy:2015efa,Boddy:2015fsa,Bartels:2017dpb,Gonzalez-Morales:2017jkx}. 
     
    \item {\boldmath $\pi^\pm\pi^\mp :$} For the annihilation channel $\chi \bar{\chi} \to \pi^\pm \pi^\mp$, the photon spectra is produced through \textit{i)} pion FSR, \textit{i.e.}, $\chi\bar{\chi}\to \pi^\pm \pi^\mp \gamma$~\cite{Coogan:2019qpu,Cirelli:2020bpc}, \textit{ii)} the main decay modes of $\pi^\pm$:
$\mu^+ \nu_\mu/ \mu^+\nu_\mu \gamma,e^+\nu_e/e^+ \nu_e \gamma$ with branching fractions taken from the PDG~\cite{Coogan:2019qpu,Cirelli:2020bpc,ParticleDataGroup:2018ovx}.

\end{itemize}

\subsubsection{Kaons}

We consider 3 different kaon final states:  $K^+K^-$, $K^0_L\, \overline{K^0_L}$ and $K^0_S\,\overline{K^0_S}$.

\begin{itemize}
  \item {\boldmath $K^+ K^-:$} Similar to charged pions, the resulting photon spectra is  a sum of the kaon FSR~\cite{Coogan:2019qpu} as well as their radiative decay.
     \textit{i)} We obtain the FSR spectrum for kaon final states, \textit{i.e.}\ $\chi\, \bar{\chi}\to K^\pm\, K^\mp\, \gamma$, using the Altarelli-Parisi (AP) approximation as described in Ref.~\cite{Coogan:2019qpu}. \textit{ii)} The relevant decay modes of $K^\pm$: $\mu^+ \nu_\mu,\pi^+\,\pi^0,\pi^0\,\mu^+\,\nu_\mu$, $\pi^0\,e^+\,\nu_e,\pi^+\,\pi^0\,\pi^0,\pi^+\,\pi^+\,\pi^-$.     
 \item {\boldmath $K^0_L\, K^0_L$} and  {\boldmath $K^0_S\, K^0_S :$ } 
  Here photons are only produced through the decay of pions and muons as described above. The relevant decay modes of $K_L^0$ we consider are 
   $\pi^+\, e^-\,\nu_e, \pi^+\, \mu^-\,\nu_\mu, \pi^0\,\pi^0\,\pi^0$ and $\pi^+ \pi^-\,\pi^0$. For $K_S^0$ we take $\pi^0\,\pi^0$  and $\pi^+\,\pi^-$.   

\end{itemize}

Again all branching fractions are taken from the PDG. The photon spectrum is computed by boosting the photon spectra from the rest frame of $\mu^+$, $\pi^0$ or $\pi^+$ to the rest frame of the decaying kaon and, eventually, from the kaon rest frame to the galactic frame~\cite{Boddy:2016hbp, Cheng:2016slx}.   

The tables containing the photon spectra from  DM annihilation into mesons are used directly when computing the photon spectra via the function  {\tt CalcSpectrum} for models in which the Lagrangian contains pions and Kaons. These must be identified by their PDG code~\cite{ParticleDataGroup:2010dbb}.

\subsection{Cosmological constraints}

Long-lived particles (LLPs) are found both in some classes of freeze-in or conversion-driven freeze-out models, as well as in scenarios in which DM is produced from decays of heavier particles after they freeze-out (\textit{i.e.}\ through the ``superWIMP'' mechanism). In such instances, different cosmological constraints can become relevant. Although in \microsix\ we do not compute these constraints, we nevertheless provide novel functionalities which can be useful in order to quantify their impact.

A first important constraint derives from Big-Bang Nucleosynthesis (BBN), since particles decaying during or after the BBN era can alter the formation of light elements in the Universe. A full study of BBN bounds requires, of course, a dedicated analysis which goes beyond the scope of the present work. However, through \microsix\ it is possible to easily obtain the abundance of LLPs and their decay modes through the routine {\tt darkOmegaN}. 
A simple way to implement the BBN constraint from hadronic energy injection as a function of the lifetime of the LLP  is to compute the quantity $B_{\rm had} E_{\rm vis} Y_{\rm LLP}$ where $B_{\rm had}$ is the hadronic fraction of the LLP, $E_{\rm vis}$ is the visible energy in the decay and $Y_{\rm LLP}$ the abundance of the LLP, 
for concrete applications see, \textit{e.g.},~\cite{Banerjee:2016uyt, Belanger:2022qxt,Belanger:2022gqc}. 

Moreover, thermally decoupled DM candidates produced through late decays of heavier particles can have larger velocities compared to their counterparts in more strongly-coupled theories and may suppress structure formation. In such scenarios small-scale structures can be washed-out due to the larger free-streaming length of the DM particles compared to the usual freeze-out picture~\cite{Hooper:2011aj}. In \microsix, we offer the possibility to calculate this free-streaming length. The latter is usually computed as an integral over time (\textit{cf e.g.} \cite{Cembranos:2005us})
\begin{equation}
\lambda_{FS} = \int_{\tau}^{t_E}  \frac{v(t)}{a(t)}  dt 
\end{equation}
where  $\tau$ is the time when the Hubble parameter equals the width of the decaying particle, $t_E$ is the time of matter-radiation equality, $v$ is the initial DM velocity $v=\frac{p/m}{\sqrt{ 1+ (p/m)^2}}$ and $a(t)$ is the cosmological scale factor, $ a = \left( s(T_0)/s(T)\right)^{\frac{1}{3}}$, where $T_0=2.725$\,K. In the code we  opt for computing the free-streaming length as an integral over T through the relation
\begin{equation}
 \lambda_{FS}=  \int
\limits_{T_2}^{T_1}  \left( 1+ \left(\frac{
a(T)m}{ a(T_1)p}\right)^2\right)^{-\frac{1}{2}}  \frac{dT}{a(T) \overline{H}(T) T }
\label{eq:freestreaming}
\end{equation}  
where $T_1$ and $T_2$ are temperatures defined by the user. The accurate computation of structure formation constraints is a highly non-trivial task, typically requiring the deployment of dedicated $N$-body simulations, for recent analyses \textit{cf e.g.} \cite{Murgia:2018now,Archidiacono:2019wdp,Hooper:2022byl}. In any case, an indicative bound $\lambda_{FS} < 0.5$ Mpc on the free-streaming length of DM particles can be obtained from Lyman-$\alpha$ forest observations, especially if a large fraction of DM comes from the superWIMP mechanism \cite{Cembranos:2005us}. More detailed treatments of the Lyman-$\alpha$ forest constraints can be found, \textit{e.g.}, in~\cite{Baumholzer:2019twf,Decant:2021mhj,Ballesteros:2020adh}.

\section{Description of routines }   
\label{sec:functions}

A complete list of \micro\ routines is 
provided in the manual contained in  the \verb|man| directory of the code, hereafter we refer to this as the on-line manual. Here, we present routines that are new or have been modified since {\tt micrOMEGAs5.0}. For completeness, we also include some of the routines that were described in~\cite{Belanger:2020gnr} and in~\cite{Alguero:2022inz}.

\subsection{Definition of the dark sectors}

For the computation of the relic density in $N$-component DM models, the particles need to be divided into {\it sectors} within each of which chemical equilibrium holds. This means that a FIMP will be in a sector of its own. An evolution equation for the abundances will be associated to each of these sectors. 
By default, the separation into DSs is defined  by the number of \verb|"~"| symbols in the beginning of the particle names. Thus, \verb|~x1| and \verb|~~x2| denote dark particles of two different sectors. Usually the sector assignment corresponds to the charge of the discrete symmetry responsible for DM stability, thus the lightest particle of each sector can be a DM candidate.
\\
\\
The default separation into sectors can be modified using the function
\\
$\bullet$~\verb|defThermalSet(n, particles_list)| \\
which moves all particles mentioned in {\tt particles\_list} to sector $n$. All particles which were assigned to sector {\it n} before this command and which do not appear in {\tt particles\_list}  are returned to their default sectors specified by the number of  \verb|~| in the beginning of their names. Particles in {\tt particles\_list} have to be separated by commas, and particle and anti-particle are automatically assigned to the same sector.  By definition, sector~$0$ is the SM bath but can also contain BSM particles with the same symmetry properties as the SM, while sector~$-1$ is used for particles that have been defined as {\it feeble}, see \verb|toFeebleList| below. 
Such particles will be ignored when solving for the relic density with the freeze-out routines. 
Sectors $n>0$, are used for all other cases. {\footnote{ Here we remind that, as already mentioned, if all particles are  in thermal equilibrium but one of them is nevertheless assigned to sector 2 which contains no other particle, two abundance equations will be solved and should give the same result as the single abundance equation, that is $Y_{\rm final}{\rm(heavier\:particles)}=0$ and $Y_{\rm final}{\rm (lightest\:particle)}=Y_{\rm final}{\rm (single\:equation)}$ of the single equation. }} 

In general, \verb|defThermalSet| can define a set which includes particles with different charges of the discrete symmetry group (different number of \verb|"~"| symbols) -- in particular the set could include $Z_2$ odd particles as well as SM particles. In this case the user must keep in mind that the abundance equations are solved for sectors $n > 0$ only. This entails that a $Z_2$-odd particle assigned to sector $0$ will not be considered as a potential DM candidate. The function returns an error code if {\tt particles\_list} contains a particle name which is not defined in the model. 

With this function it is now possible to use \micro\ for models which do not follow the usual conventions of having DS particles assigned a name that starts with a \verb|~|.
\\
\\
$\bullet$~\verb|printThermalSets()|  \\
prints the contents of all particle sets specified by \verb|defThermalSet| on the screen.
\\
\\
$\bullet$~\verb|checkTE(n, T, mode, Beps)| \\
checks the condition for chemical equilibrium  in the $n^{th}$  sector at temperature \verb|T|. If {\tt mode}=0, then both decay and  co-scattering are taken into account. If {\tt mode}=1 (2),  then only decay (co-scattering) processes are taken into account.  \verb|checkTE|  returns  the minimal value of $\Gamma/H(T)$ obtained after testing all possible subsets of particles in sector $n$. The particle assignment corresponding to the minimal value of $\Gamma/H(T)$  is printed on the screen. This value has to be $\gg X_f$ to have chemical equilibrium, when this condition is satisfied the correction to the abundance calculated assuming chemical equilibrium is approximately $\Delta Y/Y\approx X_f H/ \Gamma$. 
\\
\\
$\bullet$ \verb|toFeebleList(particle_name)| \\
assigns the particle \verb|particle_name| to the list of feebly interacting particles in sector ``-1". Feebly interacting particles can be odd or even.   
To include more than one particle, this function has to be called several times.  All  odd or even particles that are not in this list are assumed to be in thermal equilibrium with the SM at high temperatures. 
The  particles belonging to sector ``-1" will be ignored  when solving for the relic density with the  freeze-out routines.
Calling {\tt toFeebleList(NULL)} will reassign all particles to the sector they belong to according to the number of  \verb|"~"| at the beginning of their names.
\\
\\
After a reassignment of any input model parameter, after changing the  sets of particles in thermal equilibrium and after defining the set of feeble particles, for initialization of \micro\ one has to call\\
$\bullet$ \verb|sortOddParticles(txt)|\\
This routine calculates the constrained parameters of the model. It returns a non-zero error code for a wrong set of parameters, for example parameters  for which some constraint cannot be calculated. The name of the corresponding constraint is written in \verb|txt|. This routine also  defines the number of dark  sectors containing particles in chemical equilibrium, {\tt Ncdm}, and finds the name  of the lightest particle in each sector, {\tt CDM[k]} ( k=1...Ncdm), as well as its mass, {\tt McdmN[k]}. It also defines the  mass of the lightest dark particle, {\tt Mcdm}, which can either be a WIMP or a FIMP.  

\subsection{Abundances and cross sections}
\label{sec:N}

\noindent
$\bullet$~\verb|darkOmegaNTR(TR, Y, fast, Beps, &err)| \\ 
solves the abundance thermal evolution equations starting from the initial temperature {\tt TR} and returns the total $\Omega h^2$ as described in~\cite{Alguero:2022inz}. The array {\tt Y} has to contain the initial abundances at the temperature {\tt TR}. After completion, {\tt Y[k-1]} contains the abundances of sector $k$ at the temperature {\tt Tend} defined by the user\footnote{By default {\tt Tend}=$10^{-3} {\rm GeV}$, however when the decay contribution is important it is preferable to choose a smaller value such as {\tt Tend}=$10^{-8} {\rm GeV}$  which corresponds to the temperature of the Universe today.}. 
The parameter {\tt TR} is assigned to the global variable {\tt Tstart}. The parameter {\tt Beps} defines the criteria for including coannihilation channels as for the \verb|darkOmega| routine in previous versions of \micro~\cite{Belanger:2006is}. The fast = 1/0 option switches between the fast/accurate calculations. This routine also fills the array {\tt fracCDMN} which contains the  
relative contribution of each sector to the relic density, {\tt fracCDMN}$[k]=\Omega_k/\Omega$. If the model has a scale dependence $Q$, its value is fixed to  $Q=\sqrt{s}$ for each process under consideration.
\\
\\
$\bullet$~\verb|darkOmegaN(fast, Beps, &err)| \\ 
calls \verb|darkOmegaNTR| to solve the equations of the thermal evolution of abundances in the temperature interval \verb|[Tend,Tstart]|. In each sector, the function looks for the temperature $T_i$ where $Y_i(T_i)\approx Y_{eq}(T_i)$. The maximum value of $T_i$ is assigned to {\tt Tstart}. This routine returns the total value of $\Omega h^2$. The relative contribution of each sector is stored in {\tt fracCDM[i]} . If the model has a scale dependence $Q$, its value is fixed to  $Q\approx \sqrt{s_{min}}+T$ where $s_{min}$ is the minimal $s$ for each group of processes occurring in the model, \textit{e.g.}\ of the type  {\tt "ijkl"}  where $0\leq i,j,k,l \leq N$.
\\
The {\tt \&err} parameter  returns the following error code :
\begin{verbatim}
32  - no WIMP
64  - Tstart is not found, at least one of the DM component was never in 
      thermal equilibrium with SM particles.
128 - problem in solution of differential equation. The reason may be 
      that the equation is stiff because Tstart is very large.
\end{verbatim}  
The {\tt darkOmegaN} routine returns {\tt NAN} if one of these errors appears.
 
In order to calculate $\langle v\sigma \rangle$ we use the {\it simpson} program to evaluate the integrals over the scattering angle and energy of collisions. This program can return the following  error codes  
\begin{verbatim} 
1 - NAN in integrand;
2 - too deep recursion;
4 - loss of precision.
\end{verbatim}
which is  passed to $\&err$.
In general, these codes can be treated as warnings, although it can be useful to check the calculation of the problematic integrals using \textit{e.g.}\ the {\tt gdb} debugging tools. More information on this tool can be found  in the on-line manual, in Section 15. The error code {\tt err} is a binary code which can signal several problems simultaneously. 
\\
\\
For the aforementioned functions, \micro\ provides the possibility to selectively exclude part of the terms in the evolution equation. 
This is realized via the string {\tt ExcludedForNDM}, which can be assigned specific keywords. For example, the keyword \verb|"DMdecay"| excludes decay processes which contribute to the DM evolution, while the keyword \verb|"1100"| excludes $1,1 \leftrightarrow 0,0$ processes. To reset and include all channels one must use {\tt ExcludedForNDM=NULL;}.
\\
\\                       
$\bullet$~\verb|YdmNEq(T,|$\alpha${\tt)}\\
calculates the thermal equilibrium abundance  for particles of sector $\alpha$, where $\alpha$  has to be presented by a text label. For instance, {\tt YdmNeq(T,"1")}.  This can be  used as an initial condition for the abundance of a WIMP in {\tt darkOmegaNTR}. 
\\  
\\
$\bullet$~\verb|YdmN(T,| $\alpha${\tt)} \\ 
contains the abundances of DM in sector $\alpha$ as  calculated by
\verb|darkOmegaN| or \verb|darkOmegaNTR|  for  $ T \in $ {\tt [Tend,Tstart]}.
\\
\\
$\bullet$~\verb|vSigmaN(T, channel)|  \\
calculates the thermally averaged  cross section  $\langle v\sigma \rangle$  in  {\tt [pb$\cdot$c]} units. Here {\tt channel } is a  text code specifying the reaction, \textit{e.g.}\ \verb|vSigmaN(T,"1100")| for $1,1 \leftrightarrow 0,0$ processes. If {\tt channel}  starts with an exclamation mark, for example \verb|vSigmaN(T,"!1100")|, then {\tt vSigmaN} returns the results of the previous call of {\tt darkOmegaN} or {\tt darkOmegaNTR} . Otherwise  {\tt vSigmaN}  recalculates all necessary cross sections using the  parameters {\tt fast, Beps} defined in the previous call of {\tt darkOmegaN} or as defined by a call to  
\\
\\
$\bullet$~\verb|setFastBeps(fast,Beps)|.
\\
\\
If the model has a scale dependence $Q$, it will be assigned the value $Q=\sqrt{s}$ for each channel.
\\
\\
In order to find the contribution of  different processes to  {\tt vSigmaN},  one can call
\\
\noindent
$\bullet$~\verb|vSigmaNCh(T, channel, fast, Beps, &vsPb)|  which  returns an array of annihilation processes   together with their relative  contributions  to the  total annihilation  cross section.   The cross section is given by the return parameter   \verb|vsPb|  in [pb c] units. The elements of the array are  sorted according to weights and the last element has weight=0.  The structure of this array is identical to vSigmaTCh which was defined for one-DM models, see the on-line manual. The input parameter  \verb|channel| is written in text format.  The memory allocated by  outCh  can be  cleaned after usage with the command {\tt free(outCh)}. The following lines of code give an example on how to use this function: 
\\
\begin{verbatim}
aChannel*outCh=vSigmaNCh(T, "1100",  Beps, &vsPb);
for(int n=0;;n++)
{  if(outCh[n].weight==0) break;
   printf(" %.2E  %s %s -> %s %s\n", outCh[n].weight, 
   outCh[n].prtcl[0], outCh[n].prtcl[1], outCh[n].prtcl[2], outCh[n].prtcl[3]);
}
free(outCh);
\end{verbatim}

There is an option to calculate the width of any particle including the contribution from channels with different numbers of outgoing particles
\\
$\bullet$ \verb|pWidthPref(particle_name, pref)|  defines the switches for the routine \verb|pWidth| introduced in previous versions. By default \verb|pWidth|  checks the value of the {\tt useSLHAwidth} flag,  if {\tt useSLHAwidth!=0}  and  there are decay data in the loaded SLHA file, then {\tt pWidth} returns the value stored in the file. Otherwise, the widths are calculated at tree-level including only  channels with the minimal number of outgoing particles. 
\verb|pref| allows to override this prescription for a single particle. It can take the values 
\begin{itemize}
\item{}0 -  widths are calculated using processes with minimal number of outgoing particles
\item{}1 -- widths are calculated using processes with minimal and next-to-minimal number of outgoing particles excluding
                       processes with  $s$-channel resonances to avoid double-counting.
                  \item{} 2- widths are read from the SLHA file - if the SLHA file does not contain widths, they are calculated as in 0
                    \item{} 3-- widths are read from the SLHA file - if the SLHA file does not contain widths, they are calculated as in 1
                       \item{} 4 -- the default option of {\tt pWidth} is used. 
                       \end{itemize}

\noindent$\bullet$ \verb|improveCrossSection(n1,n2,n3,n4, Pcm, &cs)|\\
allows one to substitute a new cross section for a given process instead of the one calculated by
micrOMEGAs at tree-level.  Here \verb|n1,n2| are the PDG codes  for  particles in the initial state and
\verb|n3,n4| for those in the final state. \verb|Pcm| is the centre-of-mass momentum and \verb|cs| is the cross section in
[${\rm GeV}^{-2}$].  This function is called just after the calculation
of the annihilation cross section in routines that calculate the  relic density
and indirect detection. \micro\  calls this routine substituting
for the last parameter the address of the memory where the calculated tree-level cross section
{\tt cs} is
stored.
 This function is useful if, for example, the users want to include a loop-improved cross section 
calculation and/or  their own implementation of the Sommerfeld effect.
\micro\  contains  a dummy version of this routine located in 
{\tt sources/improveCS.c} which does not modify the default  cross section.
 This file also  contains some commented out example of the code for the IDM model.
 To activate this functionality the users have to write their own
version of the {\tt improveCrossSection} routine and place it  in the directory
\verb|MODEL/lib|. Then, the {\it dummy} version will be ignored.
\\
\\
$\bullet$ \verb|Sommerfeld(a,b)|\\ 
this function can be used  to calculate  the Sommerfeld enhancement for s-channel scattering processes when the mediator is either a scalar or a vector, following section~\ref{subsec:Sommerfeld} based on \cite{Iengo:2009ni}. The  {\tt improveCrossSection} routine can then be used to compute the enhanced cross section.  The arguments are defined in Eq.~\eqref{eq:Sommerfeld} 
for the Lagrangian in Eq.~\eqref{eq:Lagrangian}. 

\subsection{ Abundances and cross sections for one-component DM}

In addition to the standard {\tt darkOmega} routine for the computation of the relic density of one-component DM~\cite{Belanger:2001fz},  we have designed new functions that allow the user to provide their own cross sections. These routines can also be used to include processes with 3 or 4 bath particles in the final state as described below.
\\
\\
$\bullet$ \verb|darkOmegaExt(&Xf, vs_a, vs_sa)|\\
calculates the DM relic density $\Omega h^2$  
using annihilation cross sections  provided by external 
functions. Here  \verb|vs_a| is the  annihilation cross section in [c$\cdot$pb] as 
a function of the temperature in [GeV]  units,  while  \verb|vs_sa| 
is the semi-annihilation cross section, see the on-line manual for more details.  \verb|vs_a| is required for all models, 
while \verb|vs_sa| is relevant only for models where semi-annihilation occurs.  The user 
can  substitute {\tt NULL} for \verb|vs_sa| when semi-annihilation is not possible.
 
 {\tt darkOmegaExt}   can also be used 
if  processes other than $ 2 \to 2$ processes contribute  to DM  annihilation. In this case the appropriate annihilation or  semi-annihilation
cross sections can be calculated by {\tt vSigmaCC} and the tabulated results stored in \verb|vs_a| and \verb|vs_sa|. If the user substitutes some function which is not in tabular form, {\tt darkOmegaExt} can be  slow as it has not been optimized.

 {\tt darkOmegaExt} solves  the Runge-Kutta equation in the interval {\tt [Tstart, Tend]} 
 where Tstart is defined automatically while {\tt Tend} is defined by the user.  {\tt darkOmegaExt} takes into account  DM
asymmetry when relevant.

The routine \verb|darkOmegaExt| can only be used to calculate the  relic density of a single WIMP
particle. 

One important application of   {\tt darkOmegaExt} is that it can be used to take into account off-shell contributions  in the calculation of the relic density.
For this one has to first compute the corresponding $2\to 3$ and/or $2\to 4$ processes with the functions listed below,  create a new function   which sums the contributions of {\tt vSigmaA}, 
{\tt vSigmaPlus23} and {\tt vSigmaPlus24} and pass this function on to {\tt darkOmegaExt}.
\\
\\
\noindent
$\bullet$ \verb|vSigmaPlus23(proc22, T, &err)|\\
calculates the contribution to  $v\sigma$ for a $2\to 3$ process  associated with the  $2\to2$ process ({\tt proc22})  for which one of the outgoing particles can be off-shell.  The contribution of the on-shell $2\to2$ process is subtracted.
\\
$\bullet$ \verb|vSigmaPlus24(proc22, T, &err)|\\
which calculates the contribution to $v\sigma$ for a $2\to 4$ process  associated with the  $2\to2$ process  for which outgoing particles are off-shell. 
Here {\tt proc22} is the name of the $2\to2$ process and {\tt T} is the temperature.
For each virtual particle, the decay channel with the largest branching fraction is chosen, the result is then divided by the corresponding branching fractions.
\\
\\
When first called, both {\tt vSigmaPlus23} and {\tt vSigmaPlus24}  tabulate the cross sections of $2\to 3$ and $2\to 4$ processes respectively  
and keep results in memory for subsequent calls  to calculate integrals over $s$. A call to {\tt sortOddParticles()} cleans the  tabulated cross sections.

\subsection{Freeze-in routines}
\label{sec:routines:freeze-in} 

Several routines are provided in \micro~to compute the DM abundance in freeze-in scenarios as described in~\cite{Belanger:2018ccd}. These can be found in the file \verb|sources/freezein.c|. To accomodate multi-component DM and the possibility of several feeble  particles, the call to these routines have been slightly modified. The actual computation of the freeze-in DM abundance can be performed with the help of three functions (the equations used for the three different cases are described in~\cite{Belanger:2018ccd}):
\\
\\
$\bullet$ \verb|darkOmegaFiDecay(TR, bathParticle, feebleParticle) | \\
calculates the abundance of {\tt feebleParticle} resulting  from the  decay of {\tt bathParticle}; \verb|TR| is the reheating temperature.\\
\\
$\bullet$ \verb|Tkappa[k]| \\
is the array that defines the cut  on the Mandelstam variables {\it t} or {\it u} for processes which feature a diagram with a $t/u$-channel propagator as described in Section~\ref{sec:FI}.  {\tt k}  is the  internal particle  number. This number can be  obtained by 
\begin{verbatim}
        k=abs(pTabPos(pName))-1
\end{verbatim}
where {\tt pName} is the particle name. This array is used in  \verb|darkOmegaFi22| and  \verb|darkOmegaFi|, presented below.
\\
\\
$\bullet$ \verb|darkOmegaFi22(TR, Process, feebleParticle, &err)|  \\
calculates the  DM abundance of {\tt feebleParticle}  taking into account only the contribution of the $2\to2$ process {\tt Process}. For example \verb|"b,B ->~x1,~x1"|  for the production of DM (here \verb|~x1|) via $b\bar{b}$ scattering. This routine allows the user to extract the contribution of individual annihilation processes. \verb|TR| is the  reheating temperature. 
{\tt err} is the  returned error code, see \cite{Belanger:2018ccd} for the meaning of different error codes.
\\
\\
$\bullet$ \verb|darkOmegaFi(TR,feebleParticle,&err)|  \\
calculates the  DM abundance after summing over all 
$2\to2$ processes involving  particles in the bath ${\cal B}$ in the initial state  and at least one particle \verb|feebleParticle|  in the final state. The routine checks the decay modes of all bath particles and if one of them has no decay modes into two other bath particles, the $2\to2$ processes involving this particle are removed from the summation and instead the contribution to the DM abundance computed from the routine \verb|darkOmegaFiDecay| is included in the sum.  This is done to avoid appearance of poles in the corresponding $2\to2$ cross section. For such models, we recommend the user to compute individual $2\to2$ contributions with the function \verb|darkOmegaFi22| described above. 
\verb|TR| is the reheating temperature and   
{\tt err} is the returned error code, \verb|err=1| if feeble particles have not been defined. 
\\
\\
$\bullet$ \verb|printChannelsFi(cut,prcnt,filename)|  \\
writes into the file \verb|filename| the contribution of different channels to $\Omega h^2$. The \verb|cut| parameter specifies the lowest relative contribution to be printed.   If
\verb|prcnt| $\neq 0$, the contributions are given in percent format.   
The routine \verb|darkOmegaFi| fills  the array \verb|omegaFiCh| which contains the contribution of different channels ($2\rightarrow 2$ or $1\rightarrow 2$) to $\Omega h^2$.
{\tt omegaFiCh[i].weight} specifies the relative weight of the $i$-th channel, while
{\tt omegaFiCh[i].prtcl[j]} $(j=0,\cdots 4)$ defines the particle names for the $i$-th channel.
The last record in the array  \verb|omegaFiCh| has zero weight and {\tt NULL} particle name.\\

One can check the  temperature evolution of the abundances generated by the last three routines by calling  the function {\tt YFi(T)} in the interval $T\in [ {\rm Tend, Tstart}]$, where {\tt Tstart=TR} is set internally by the routines. This has no influence on the relic density calculation but is used for the {\tt displayPlot} function.

If no particle has been declared as being feebly interacting, the freeze-out routines \verb|darkOmega|,   \verb|darkOmega2|~\cite{Belanger:2014vza}, \verb|darkOmegaN| and {\tt darkOmegaNTR} will work exactly as described above or in previous versions of \micro. A non-empty \verb|feeblelist|, however, will affect these routines since \micro~will exclude all
 the particles in this list from the computation of the relic density via freeze-out.
To compute the relic density of particles in  \verb|feeblelist| one has to use the {\tt darkOmegaFi} function (or the other routines described in this section) for each of these particles. Alternatively, the relic density of feeble particles can be computed using the {\tt darkOmegaNTR} routines described in section~\ref{sec:N} by setting the initial abundance of these particles to zero and removing them from \verb|feeblelist|. Note that the result can differ from  the one obtained through the {\tt darkOmegaFi} routine since the $N$-component functions assume Maxwell-Boltzmann statistics. \footnote{In previous versions of \micro\ the relic density of {\tt darkOmegaFi} was rescaled when it was not the lightest particle in the DS to take into account the fact that the FIMP will eventually decay into the DM. In this version this rescaling has to be done by the user.} 

\subsection{Thermodynamics}

The effective number of degrees of freedom can   be accessed through the functions:
\\
\\
$\bullet$ \verb|gEff(T)|\\
 which returns the effective number of degrees of freedom for the energy density of radiation at a bath temperature \verb|T|, including SM particles only.
\\
\\
$\bullet$ \verb|hEff(T)|\\ which returns the effective number of degrees of freedom for the entropy density of radiation at a bath temperature
\verb|T|, only including SM particles.

The  default tables for  $h_{\eff}, g_{\eff}$ correspond to the ones in Ref.~\cite{Drees:2015exa} and can be found in the file \verb|Data/hgEff/DHS.thg|.
\\
\\
These default tables can be changed using  \\
\noindent$\bullet$ \verb|loadHeffGeff(char*fname)|\\
that  reads  the file {\tt fname} located in the directory {\tt Data/hgEff}. This file should 
 contain  3 columns for $T$, $h_{\eff}(T)$ and $g_{\eff}(T)$. 
A positive  return value corresponds to the number of lines in the table. A negative return value indicates the line which creates a problem (\textit{e.g.}\ wrong format), while the routine returns zero when the file \verb|fname| cannot be opened.

The  directory  {\tt Data/hgEff}  also contains solutions described in Ref.~\cite{Laine:2015kra}, {\tt LM.thg},  and in Ref.~ \cite{Hindmarsh:2005ix},  {\tt HP\_B.thg, HP\_C.thg},  as well as the tables used in {\tt DarkSUSY}, {\tt GG.thg}. The latter were used as default in  previous versions of \micro\ and do not include the contribution of the Higgs boson. 
\\
\\
$\bullet$ \verb|hEffLnDiff(T)|\\ returns the derivative of $h_{\eff}$ with respect to the
 $\log$ of the bath temperature, $\frac{d\log(h_{\eff}(T))}{d\log(T)}$.
\\
\\
$\bullet$ \verb|Hubble(T)|\\ returns the Hubble expansion rate in
{\tt GeV} units   at a bath temperature \verb|T|[GeV].
\\
\\
$\bullet$ \verb|HubbleTime(T1,T2) |\\
 calculates  the time interval in  seconds  during which the  temperature of the Universe
decreases from {\tt T1[GeV]}  to {\tt T2[GeV]}, as obtained from Eq.~\eqref{eq:hubbletime}. 
\\
\\
$\bullet$ \verb|T2_73K|\\
 gives  the  current temperature T= 2.725K in GeV units. 
\\
\\
$\bullet$ \verb|freeStreaming(p/m, T1,T2)|\\
calculates  the free-streaming length in Mpc units for a freely-propagating particle between the temperatures
{\tt T1[GeV]} and  {\tt T2}, by employing Eq.\eqref{eq:freestreaming}. The parameter {\tt p/m} 
characterizes the initial velocity of the particle $v=\frac{p/m}{\sqrt{ 1+ (p/m)^2}}$.

\subsection{Direct and Indirect  detection  }

The micrOMEGAs' routines to compute the  cross sections for elastic scattering of DM on nucleons can be used for N-component DM as well, the output is $\sigma_{\chi_i p/n}$ for either spin-dependent or spin-independent interactions.
In addition, the following routine based on a recasting of several experimental limits~\cite{Belanger:2020gnr}  allows to determine the level of exclusion for a given model,

\noindent
$\bullet$ \verb|DD_pval|(\verb|expCode|, $f_v$, {\tt \&expName})\\
calculates the  value $\alpha= 1-C.L.$  for the  model under consideration including the contribution of the N-components.
The return value 0.1 corresponds to a 90\% exclusion.   
 The {\tt expCode} parameter can be any of the codes  \verb|LZ5Tmedian,XENON1T_2018,DarkSide_2018|, \verb|CRESST_2019,PICO_2019| or their combination concatenated with the symbol
$\mid$. There is also a predefined parameter that  currently combines these experiments 
\begin{verbatim}
  AllDDexp=LZ5Tmedian|XENON1T_2018|DarkSide_2018|PICO_2019|CRESST_2019;
\end{verbatim}

The parameter {\tt char* expName}  is used to indicate the   experiment that  provides
the best exclusion among those specified in  {\tt expCode}.   The function {\tt DD\_pval } calculates the
exclusion for each experiment  independently, returns the smallest $\alpha$, and assigns the 
name of the corresponding experiment to  {\tt expName}  if it is not {\tt NULL}.  

The $f_v$ parameter specifies the DM velocity distribution in the detector frame. For
example, one can use   {\tt Maxwell} or {\tt SHMpp} which are
included in \micro, otherwise the user can define another distribution, see the on-line manual.  
 The units  are $km/s$ for $v$ and $s/km$ for  $f_v(v)$. {\tt DD\_pval} implicitly depends on the 
global parameter {\tt rhoDM} which specifies  the DM  local density respectively.

For  Xenon1T one can choose  between  three recasting methods, $p_{\eff}^{q}$ with $q=0,1,2$, 
see Ref.~\cite{Belanger:2020gnr}. The  flag {\tt Xe1TnEvents=q} allows to choose the  corresponding recasting,  otherwise and by default the code uses  $p_{\eff}^1$. The three approaches agree  within 5\%. For PICO-60, the user can choose between the recasting based on Feldman-Cousins statistics, {\tt PICO60Flag=0} 
which is the default value, or the one based on  Neyman one-side belt exclusion, {\tt PICO60Flag} =1.\\ 

\noindent
$\bullet$  \verb|DD_factor|(\verb|expCode|, $\alpha$, $f_v$,\&expName)\\     
returns the overall factor which should be applied to all the cross sections, $\sigma^{\rm SI}_{\chi_i p} ,\sigma^{\rm SI}_{\chi_i n}$ or $\sigma^{\rm SD}_{\chi_i p}, \sigma^{\rm SD}_{\chi_i n}$ to reach  the exclusion level $\alpha$.  
All parameters are the same as in {\tt DD\_pval} above. This is intended to give the user an indication on how strongly a given model is excluded.\\

For DM indirect detection, several tables that contain the 
spectra of stable particles from DM pair annihilation into two particle final states are provided. 

\noindent
$\bullet$   \verb|SpectraFlag| \\
is a switch to choose the tables that contain the spectra  for $\gamma,e^+,\bar{p},\nu$ from DM pair annihilation into two particle final states. 
{\tt SpectraFlag=0} corresponds to the tables obtained with  Pythia-6 \cite{Belanger:2010gh}  for  DM mass in the range  $2{\rm  GeV} - 5{\rm TeV}$ for ,$e^+,\bar{p},\nu$. 
For the photon channel the tables  have  been extended to cover the range $110 {\rm MeV} - 5{\rm TeV}$ as described in section~\ref{sec:ID}.
These tables include also the spectra for polarized W's and Z's.
This Flag has to be set before calling the functions {\tt calcSpectrum} and {\tt basicSpectra}. 
{\tt SpectraFlag=1} corresponds to the spectra generated by Pythia-8 \cite{Amoroso:2018qga, Jueid:2022qjg} for DM mass in the range  $5{\rm  GeV} - 5{\rm TeV}$, here polarization is not taken into account. 
The  QCD uncertainties on the spectra discussed in Ref.~\cite{Amoroso:2018qga, Jueid:2022qjg}   are not yet implemented in  {\tt micrOMEGAs}. 
{\tt SpectraFlag=2} corresponds to the 
 spectra generated with PPPC [72, 73] for DM mass larger than 5GeV. The spectra take into account polarized W and Z,  as well as
polarized leptons,  the latter are however not supported
in the current version of {\tt micrOMEGAs}.\\

\noindent
{\bf Graphic representation of  DM evolution} \\

The routine {\tt displayPlot} allows to display on the screen many quantities computed with  {\tt micrOMEGAs}, see section 15.5 of the on-line manual for a complete description.  For example the following command
\begin{verbatim}
displayPlot("Y","T",T1,T2,scale,np,"Yeq1",0,YdmNEq,"1","Y1",0,YdmN, "1");
\end{verbatim}
will display $ \overline{Y_1}(T)\equiv$ {\tt YdmNEq(T,"1")} and $Y_1\equiv${\tt YdmN(T,"1")} in the temperature interval $[T_1,T_2]$. Here \verb|scale=0(1)| sets a linear (log) scale and {\tt np} is the number of curves to be plotted (here {\tt np=2}).

\section{Examples}
\label{sec:examples}

A sample main file is provided in the two singlet model with $Z_5$ symmetry,  it can be found in  \verb|Z5M/CPC_fig.c| together with a sample data file {\tt fimpf.par}. This file  will output the data necessary to prepare  Fig.~\ref{fig:om2omN} (right).

The output that will be produced on the screen gives the result of the two DM relic density calculation, here \verb|~x1|  is a FIMP while \verb|~~x2| is a WIMP. The results compare the output of \verb|darkOmegaNTR|, denoted as {\tt Omega1 (N) }  below, with the one from \verb|darkOmega| for the WIMP and from \verb|darkOmegaFI| for \verb|~x1|. The channels that contribute to freeze-in are also listed.

\begin{verbatim}
DM candidate is '~x1' with spin=0/2 mass=1.00E+02

DM candidate is '~~x2' with spin=0/2 mass=3.50E+02

=== MASSES OF HIGGS AND ODD PARTICLES: ===
Higgs masses and widths
      h   125.00 3.07E-03

Masses of odd sector Particles:
~x1      : Mdm1    = 100.000 || ~~x2     : Mdm2    = 350.000 || 

==== Calculation of relic density =====
  Detected types of  reactions:
Decays of Odd particles are included in equations
      2  2  <-> 0 0
      2  2  <-> 1 1
      2  2  <-> 1 0
      2  2  <-> 2 1
      2  1  <-> 1 0
      2  1  <-> 1 1
      2  1  <-> 2 0
      2  0  <-> 1 1
      1  1  <-> 0 0
New Tab ok
Omega1 (N)=1.88E-01
Omega2 (N)=2.91E-13
Omega_FI=1.52E-01
# Channels which contribute to omega h^2 via freeze-in
 6.535E-01  ~~x2 -> ~x1, ~x1               
 1.634E-01  W-, W+ ->  ~x1,~X1             
 8.855E-02  h, h ->  ~x1,~X1               
 7.622E-02  Z, Z ->  ~x1,~X1               
 1.818E-02  T, t ->  ~x1,~X1               
 1.249E-04  G, G ->  ~x1,~X1               
 8.636E-05  B, b ->  ~x1,~X1               
 1.105E-05  L, l ->  ~x1,~X1               
 4.182E-06  C, c ->  ~x1,~X1               
 1.788E-06  A, A ->  ~x1,~X1               
 4.201E-07  S, s ->  ~x1,~X1               
 3.912E-08  M, m ->  ~x1,~X1               
 1.050E-09  U, u ->  ~x1,~X1               
 1.050E-09  D, d ->  ~x1,~X1               
\end{verbatim}

To get the abundance for the FIMP including the Maxwell-Boltzmann statistics that is shown in Fig.~\ref{fig:om2omN}, one has to uncomment the first line of the file \verb|sources/freezein.c| that contains \verb|#define NOSTATISTICS|, recompile {\tt micrOMEGAs} and rerun the code.

\section{Conclusion}
\label{sec:conclusions}

In this paper, we presented a new version of the \micro\ DM code that computes the abundances of N-component DM including models with multiple WIMPs, FIMPs or a combination of WIMP and FIMP. The Boltzmann equations have also been generalised  to include the co-scattering mechanism as well as the decays of unstable DS particles. The new routines are compatible with those of previous versions in the case of one-component DM (WIMP or FIMP) or two-component WIMP DM.

We have, moreover, improved the freeze-in routine introduced in a previous version~\cite{Belanger:2018ccd} to account for the effects of the thermal masses of particles exchanged in $t/u$-channel processes by introducing an integration cut. We demonstrated the equivalence between this cut and a thermal mass by an explicit calculation of a standard model process with $t$-channel exchange of a vector particle. 

The routines for predicting the rates relevant for direct or indirect detection take into account all components, such that
constraints on the combined signal of all DM components can be imposed. 
We have also extended the tables for DM annihilation-induced photon production to masses as low as 110~MeV. To this end, we have replaced previous tables for DM annihilation into electrons and muons, and we have added new tables for DM annihilation into pairs of light mesons (pions and kaons). 

The new version, \microsix, is available at \url{https://zenodo.org/doi/10.5281/zenodo.10462240}~\cite{alguero_2024_10462241} 
and can readily be used for DM phenomenology beyond the standard WIMP paradigm.

\section*{Acknowledgements}

We thank C.~Yaguna and \'O.~Zapata for the implementation of the two singlets $Z_5$ model and three singlets $Z_7$ model. AG aknowledges useful discussions with G.~Arcadi, F.~Costa and O.~Lebedev. 

GB and AG acknowledge support by Institut Pascal at Université Paris-Saclay during the Paris-Saclay Astroparticle Symposium 2023, with the support of the P2IO Laboratory of Excellence (program “Investissements d’avenir” ANR-11-IDEX-0003-01 Paris-Saclay and ANR-10-LABX-0038), the P2I axis of the Graduate School of Physics of Université Paris-Saclay, as well as IJCLab, CEA, IAS, OSUPS, and the IN2P3 master project UCMN. 
GA and SK were supported in part by the IN2P3 master project “Th\'eorie – BSMGA”.
SC acknowledges support by the Future Leader Fellowship ‘DARKMAP’ and the hospitality of LAPTh Annecy where part of this work was done. 
The work of AP was carried out within the scientific program “Particle Physics and Cosmology” of the Russian National Center for Physics and Mathematics. 
GB and AP thank NYU Abu Dhabi  for their hospitality during the time this work was finalised. 




\providecommand{\href}[2]{#2}\begingroup\raggedright\endgroup

\end{document}